# Constrained Generalization for Data Anonymization: A Systematic Search Based Approach

BIJIT HORE,
University of California, Irvine
RAVI CHANDRA JAMMALAMADAKA,
University of California, Irvine
SHARAD MEHROTRA
University of California, Irvine
AMEDEO D'ASCANIO
University of Bologna, Italy

---

Data *generalization* is a powerful technique for *sanitizing* multi-attribute data for publication. In a multidimensional model, a subset of attributes called the *quasi-identifiers* (QI) are used to define the space and a generalization scheme corresponds to a partitioning of the data space. The process of sanitization can be modeled as a constrained optimization problem where the information loss metric is to be minimized while ensuring that the privacy criteria are enforced. The privacy requirements translate into constraints on the partitions (bins), like minimum occupancy constraints for $k$-anonymity, value diversity constraint for $l$-diversity etc. Most algorithms proposed till date use some greedy search heuristic to search for a locally optimal generalization scheme. The performance of such algorithms degrade rapidly as the constraints are made more complex and/or numerous. To address this issue, in this paper we develop a complete enumeration based systematic search framework that searches for the globally optimal generalization scheme amongst all feasible candidates. We employ a novel enumeration technique that eliminates duplicates and develop effective pruning heuristics that cut down the solution space in order to make the search tractable. Our scheme is versatile enough to accommodate multiple constraints and information loss functions satisfying a set of generic properties (that are usually satisfied by most metrics proposed in literature). Additionally, our approach allows the user to specify various stopping criteria and can give a bound on the approximation factor achieved by any candidate solution. Finally, we carry out extensive experimentation whose results illustrate the power of our algorithm and its advantage over other competing approaches.

---

Author's address: Bijit Hore, Department of Information and Computer Science, University of California, Irvine CA

## 1. INTRODUCTION (JANUARY 2011)

Data *sanitization* for publishing has been a very active area of research over the past decade. When the dataset contains information about individuals, sanitization first removes explicit identifiers like name, street address, social-security number etc. and then, carries out data transformations to reduce risk of linking records to external data as well as the risk of disclosing sensitive information. The objective is to determine the set of data transformations that retain maximum amount of information, while implementing all the privacy constraints. Data sanitization has turned out to be a challenging problem due to the presence of numerous inference channels that an adversary can exploit to disclose sensitive values and discover identities of people. Research has explored mainly the following two directions in this domain:

(1) One set of research efforts have been towards identifying the appropriate privacy criteria for data publishing. Over the years researchers have achieved a better understanding of privacy and its implications, and as a result, a variety of privacy formalizations have emerged. For instance, $k$-anonymity, $l$-diversity, $t$-closeness, $(\alpha, \beta)$-privacy and $\epsilon$-differential privacy are some of the important privacy definitions that have been proposed. We will describe some of these in greater detail later.

(2) The second is towards developing efficient sanitization algorithms. The sanitization process is often posed as a constrained optimization problem. Most of these formulations are NP-Hard and algorithms use greedy heuristics for search that lead to locally optimal solutions. A couple of notable exceptions are [Bayardo and Agrawal 2005; **?**] that develop a complete search strategy for determining the globally optimal solution to the sanitization problem. One of these papers is our own [**?**] which is a precursor of the current paper.

Our paper falls in this second class, i.e., it is about sanitization algorithms, specifically about generalization. As noted above, most popular sanitization algorithms use some greedy heuristic for search and may or may not produce a high quality solution. Typically, the quality of solutions generated by such algorithms degrade rapidly as the constraints become more complex and/or numerous. In this paper, we address this shortcoming in context of the generalization approach to sanitization of multidimensional data. Specifically, we develop an exhaustive search mechanism that enables the user to control the progress of the search algorithm. It allows him to determine how the space of feasible solutions is searched and to specify when to stop. We employ a novel enumeration technique that generates all instances of space partitioning using a specified set of attribute values to split the space along each dimension. More importantly, it avoids duplicate solutions in different branches of the enumeration tree and uses effective pruning heuristics to cut down the solution space and make the search tractable. Our scheme is versatile enough to accommodate a variety of constraints and information loss functions. As mentioned above, our algorithm allows the user to specify a stopping criteria and can give a bound on the approximation factor achieved by any candidate solution.

**Background on data sanitization:** Research has explored basically three broad approaches to sanitization for data publishing - *generalization* where attribute values are generalized, *suppression* where attribute values might be completely suppressed (this is an extreme form of generalization) and *perturbation* where a noise is added in order to mask the true value of the attribute. However, most recent work has focussed on the generalization approach. For instance, consider the well known problem of *k-anonymizing* a dataset containing data pertaining to individuals. In a multidimensional model, a subset of attributes called the *quasi-identifiers* (QI) are used to define the space and a generalization scheme corresponds to a partitioning of the data space. After partitioning, the values of the QI attributes are generalized for all elements in each bin to reflect the extent of the bin. Typically, QIs are a set of those attributes that may be linked with other databases to reveal an individual's identity. For example, if the data has the following attributes - *age, zip-code, race* and *illness*, the first three attributes taken together may be designated as the QI set. Let the privacy criterion be 3-anonymity and the objective be to minimize the information loss (measured by some appropriate cost function). After sanitizing the data to meet the privacy constraints, attribute values of records will get generalized, i.e., made less specific. For example, an individual with $age = 33$, $zip = 92301$ and $race =$ "*Indian*" maybe replaced by $age \in [30, 35]$, $zip = 923 * *$ and $race =$ "*asian*", which is basically the extent of the bin it belongs to. Since all records in a bin have the same QI attribute value (i.e., indistinguishable along the QI dimensions), they are said to form an *equivalence class*.

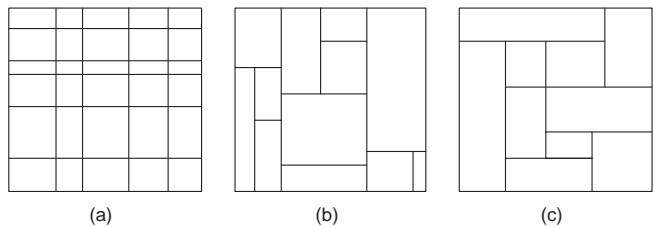

Fig. 1. Example of partitioning schemes (a) Grid-based (global recoding), (b) Hierarchical (local recoding), (c) Generic (local recoding).

Given the class of sanitization mechanism to use, there is a space of all possible data representations that satisfy the privacy criteria, i.e., the *feasible set*. In general, the sanitization algorithm consists of three components: (i) Defining the search space of all possible sanitizations. (ii) Specifying the optimization metrics and privacy constraints. When an information-loss metric is used, the problem is posed as a minimization problem, and when the metric estimates the utility of the sanitized data, the problem is posed as a maximization one. (iii) Constructing the search algorithm that searches the space of all feasible solutions.

**Existing Generalization algorithms**: Many generalization algorithms use a taxonomy based partitioning technique where each attribute has a taxonomy defined on its domain[1]. The algorithm selects a generalized representation (node in the taxonomy tree) along each dimension which then jointly determine a partitioning of the whole multidimensional space. Algorithms that consider each attribute independently from the other attributes are referred to as *global recoding* schemes. For example, global recoding of a 2-dimensional dataset may look like the one shown in figure 1 (a). We will refer to these as *gridding schemes* since they generate grid-like partitions where each splitting plane goes all the way across. Algorithms proposed in [Sweeney 2002; Samarati and Sweeney ; Bayardo and Agrawal 2005; Iyengar 2002; LeFevre et al. 2005; Fung et al. 2005; Aggarwal and Yu 2004] fall in this category. One can minimize the information loss much further by considering *local recoding* schemes (*multidimensional generalization*) schemes. These partitioning schemes allow for greater flexibility in determining the extents of a bin in a region depending upon local distribution of data. Therefore, unlike in grids, intervals along a dimension can differ depending upon the value of other attributes. Figure 1 (b) and (c) show two instances of such partitioning. It is easy to see that gridding schemes form a subset of multidimensional schemes.

---

[1]Such taxonomies may be available beforehand from domain experts for categorical attributes or created using interval creation or discretization as in the case of numeric attributes.

Irrespective of whether global or local recoding is used, the optimal data generalization almost certainly turns out to be a computationally challenging (NP-hard) problem. Most algorithms use some greedy heuristic for search that lead to locally optimal solutions. A notable exception was [Bayardo and Agrawal 2005] where the authors developed a complete search strategy for finding the optimal global recoding scheme for a multidimensional dataset. However, Lefevre et al. demonstrated in [LeFevre et al. 2006] that if one considers the space of all local recoding schemes representable by a kd-tree, then using a simple greedy heuristics one is able to quickly arrive at generalizations that have smaller cost (information loss) than even the optimal gridding schemes computed in [Bayardo and Agrawal 2005]. In fact, the authors in [LeFevre et al. 2006] simply adapted the kd-tree [Friedman et al. 1977] algorithm to greedily select the next feasible splitting plane (i.e., one that does not violate the privacy constraint) that leads to the steepest decrease in cost.

**Salient features of the enumeration based approach**: There are numerous advantages of our technique over greedy techniques such as [LeFevre et al. 2006]. First, unlike most algorithms, our technique is incremental, wherein we can achieve better results by allowing the search algorithm to run longer. In contrast, greedy algorithms terminate quickly but can produce only a single answer which achieves a local minima. In fact, our algorithm can take the solution generated by a greedy algorithm as a seed solution and improve upon it unless it happens to be global minima. Further, unlike most polynomial time algorithms, the enumeration based one can provide reasonable approximation factors to any candidate solution. While the greedy algorithm in [LeFevre et al. 2006] can also determine a bound on the quality of solution, it works only for the specific information loss metric that is used in the paper. In the enumeration based approach, since we always maintain a lower bound to the globally optimum solution (which gets tighter as the algorithm progresses) we can always determine a bound of the ratio of the cost achieved by a partitioning scheme to the optimum cost. This works for all metrics and constraints. These advantages are further elaborated below.

*Versatility of search*: Our search strategy is not tied to a specific optimization problem, but is applicable to a much larger class of problem settings. Other enhanced techniques such as *entropy l-diversity, (c,l)-recursive diversity* [Machanava-jjhala et al. 2007], *t-closeness* [Li et al. 2007] and more recently $\epsilon-privacy$ [Machanava-jjhala et al. 2009] have been proposed to address the inadequacies of the previous criteria and their effectiveness. Our algorithm can work with these more robust definitions as well. Even for the *k*-anonymization problem, an user may want to impose additional privacy constraints, such as *length restrictions* of anonymity groups along certain attribute dimensions. For instance, the following restriction could be imposed "In the released data set, no individual's salary should be specified to an interval of length smaller than 10K" and/or "Age of any individual should not be specified to an interval less than 5 years" etc. Our solution approach can be seamlessly extended to these and many other new problem settings where the cost functions and problem constraints satisfy some generic monotonicity properties[2](to be discussed in section 3). Indeed, we will show the performance characteristics of our algorithm for a variety of such problem settings in the experiments section.

*Progressive improvement in solution quality*: Our priority queue based algorithm, at all time maintains a lower bound to the true optimum (global minimum) cost. This feature allows one to stop the algorithm as soon as a solution with cost within an user-specified approximation factor is reached. As the search progresses, better approximations to the optimum solution are generated by simultaneously lowering the solution costs and tightening the lower bound estimates. Of course, if let to run till completion, the algorithm does a complete search of the solution space and guarantees finding all solutions with optimum cost.

*Handling of categorical attribute*: In many cases attributes can be *categorical*, as they do not have an implicit order defined on their domains. Using a taxonomy tree it is possible to define a partial order on the domains of this kind of attribute. Our approach is also designed to handle such *categorical attribute.*

*Incorporation of search heuristics*: The priority based approach has an added flexibility, it allows for variety of application driven search strategies/heuristics to be incorporated into the algorithm to enable one to find a good solution early. An early detection of a good solution benefits in two ways: First, it helps establish a good lower bound for pruning purposes. Second, it helps reach desired level of approximation quickly. We introduce a few of these heuristics in section 4.2 and report our findings in the experimental section.

The overall techniques we develop provides for a powerful approach to exploring the enormous space of hierarchical partitioning based solutions for a variety of anonymization and privacy constraints.

This paper extends the previous work done in [?] by the authors in a major way. In particular, the new part consists of the following - (i) the basic enumeration and pruning algorithms have been described in detail; (ii) The proof of the completeness and uniqueness properties of the enumeration algorithm has been provided; (iii) a new partition enumeration algorithm for heterogeneous spaces (mixed numeric and categorical attributes) is proposed; (iv) extensions to the basic algorithm and pruning heuristics to incorporate an important new privacy criterion called *epsilon*-privacy [Machanavajjhala et al. 2009] has been given; (v) many more experiments with new privacy constraints and utility metrics have been carried out.

---

[2]Most popular cost measures and constraints occurring in optimization problems related to anonymization satisfy these properties.

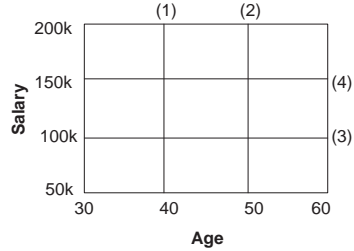

Fig. 2.   Finest partition of the space

## 2.   ENUMERATING PARTITIONS

From here onwards, for notational convenience we will use the term "partition" and "partitioning" interchangeably to refer to a complete partitioning scheme of the space. To refer to an undivided block of space within a partitioning scheme, we will use the term "partition block" or simply the term "block".In this section we describe the enumeration algorithm for hierarchical partitions of the space.

Let us consider a $d$-dimensional space where each dimension corresponds to one attribute of the data set. For instance, the figure 2 shows a 2-dimensional space where one dimension is $Age$ and the other $Salary$ with domains [30yrs, 60yrs] and [50K, 200K] respectively. Let the split set comprise of the four splits: (1):Age=40, (2):Age = 50, (3):Salary = 100K and (4):Salary = 150K. The figure 2 shows the finest level of partition (where all splits are used and all of them cut across the whole space). A hierarchical partition of a d-dimensional space consists of a set of disjoint d-dimensional hyper-rectangles that cover the entire space. Such a partition can be generated by recursively splitting the space (we will give formal definition a little later). Such partitionings can be represented by binary trees, see figure 3 for example.

The objective of our enumeration algorithm is to generate the set of all possible hierarchical partitions. The difficulty arises in trying to generate it in duplicate free manner. To see the problem of duplicates, consider the figure 3 again. The binary tree in the figure could have been created by the following sequence of splits: (3), (1), (4), (2) and (2), where the integer within '(' and ')' is the split-id and the integer within '[' and ']' is the timestamp at which the node was split. Here we follow the rule that nodes earlier in DFS order should be split before those that come later (if they are split at all that is). Now, notice that this is not the only way this partition could have been generated, it could have also been generated by the following sequence of splits: (4), (3), (1), (2) and (2). Though the final partitions are identical, the binary trees resulting from these two sequences are structurally very different.

The compelling reasons to avoid duplicate generation are the following: (i) Our approach, as in any optimization algorithm, will enumerate solutions and choose the best one based on cost. If duplicates are generated, the number of elements generated explodes. For instance take the simple case of one dimensional partitions with N split values. The true number of distinct partitions is $2^N$, whereas if we did a blind enumeration along each branch of the enumeration tree, we will end up with $N!$ (which is roughly $\sqrt{(2n + 1/3)\pi} n^n e^{-n}$ by Stirling's approximation) total partitions, which is exponentially larger than $2^N$. While it is straightforward to eliminate duplicates for the set enumeration problem by imposing an order on the splits, it is much more complex to eliminate duplicates in the case of general hierarchical partitions in arbitrary dimensional spaces. (ii) The second reason to avoid duplicates is also motivated by the nature of branch and bound solution approach we take, whose performance depends on the effectiveness of pruning. Duplicate generation severely degrades the prunability of branches, therefore practically rendering such an approach useless since many (all) good solutions will appear in all branches of the enumeration tree.

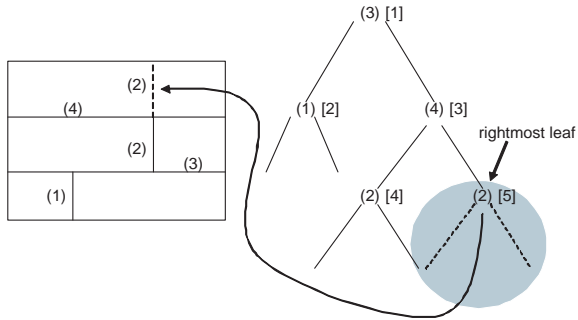

Fig. 3.   A sequence of splits generating a partition

We now give some useful definitions before presenting our partition enumeration problem.

## 2.1   The Space of Hierarchical Partitions

Let us assume for ease of visualization that each attribute is numeric, i.e., coming from a totally ordered domain and has been appropriately discretized. (Extension of our methods to categorical attributes are described in the next subsection). Then the *split set* is defined as follows.

**Definition** 1. **(Split set):**  *The set of all (attribute, value) pairs at which the space is allowed to be partitioned, is called the split set and is denoted by $S$.*

**Definition** 2. **(Split ordering):**  *We impose an absolute order on the elements of the split set defined above. Therefore each ordered pair: $(attribute, value)$ denotes an unique split from this set. A lower id of a split denotes a higher priority[3]. All split values of an attribute belong to a single* **group** *labelled by that attribute (i.e. groups $S_1, \ldots, S_m$ are the set of splits belonging to attributes $A_1, \ldots, A_m$ respectively, where $S = S_1 \cup S_2 \cup \ldots S_m$). In a geometric interpretation, splits in a group correspond to parallel planes in the space (In figure 2 splits (1) and (2) belong to one group and splits (3) and (4) belong to another group).*

---

[3]For attributes with ordered domains, we assign ids to the splits sequentially, where splits corresponding to a higher values of the attribute get higher ids. Though any random sequence can be assigned as well.

If a common split runs across some rectangular subspace, we say it is a *cut* across that subspace. A cut is formally defined as follows:

**Definition** 3. **(Cut of a subspace):** *A split* **s** *is said to form a* **s-cut of a subspace** *H if the union of* **s***-cuts across one or more of the partition blocks constituting H also runs across the whole subspace (i.e. divides H into two disjoint parts).*

For example, in figure 3, split (3) is denoted as the (3)-cut across the whole space and split (4) is the cut of the subspace denoted by the right child of the root.

Now, a hierarchical partition[4] of the (multidimensional) domain space of the data is defined as follows:

**Definition** 4. *(***Hierarchical Space Partition***): Given a multidimensional space H and a set of candidate splits S such that each $s \in S$ is a potential s-cut of H, then a hierarchical partition of the space is achieved by either leaving H undivided or by choosing a cut $s_0$ of H and then recursively partitioning the two resultant subspaces $H_1$ and $H_2$ hierarchically using splits from the set $S - \{s_0\}$.*

A hierarchical partition can be represented by a labelled binary tree which we call the **partition tree**. We note that a partition tree belongs to a class of trees called the **kd-tree**[5] [Bentley ], which is a popular data structure to represent multidimensional partitioning of a space. We represent each partition in our solution space by a corresponding partition tree as defined below.

**Definition** 5. **(Partition tree):** *A partition tree is a labeled binary tree that represents a hierarchical partition of the space. Each internal node of the tree denotes a subspace that is further partitioned into two or more subspaces. A leaf denotes an undivided subspace (or a partition block). Each internal node is associated with a split-id denoting the split that was used to cut this subspace into two. Left and right side for a split s correspond to the side with lower split ids and higher split ids than s in its group respectively. Therefore leftchild and righchild are well defined for each node in a partition tree. The partition tree is identical to the kd-tree representation of a space partition [Bentley ].*

A new partition tree is generated from a parent tree by splitting one of its leaf nodes (i.e. a partition block), which then becomes an internal node with two newly generated leaves as it children. See figure 3 which shows how addition of a new split to an existing partition is reflected in the corresponding partition tree.
**Generating partition trees:** Our enumeration algorithm (to be described shortly) generates partitions incrementally, starting with the undivided space and splitting one partition block at a time. The starting state is denoted by a partition tree which has single node, the *root*. At this point the partition of space consists of just one block, to which all data belongs. Then, the first split, say $s_i$ is chosen at the root which divides the space into two new blocks. The new partition is represented by a two level binary tree and the two leaves denote the *leftchild* and *rightchild* of

---

the root. A split applied to a node $n$ in the partition tree is denoted by the label "$(s_i)$" beside $n$ in the tree. The remaining set of splits $(S - \{s_i\})$ get propagated to the newly formed leaf nodes in the following manner: All the splits that are not in the same group as $s_i$ (i.e., do not belong to the same attribute as $s_i$) are also available at each child. Out of the remaining splits in $group(s_i)$, the ones to the "left" of $s_i$ go to the left child and the ones to its "right" go to the right child. These propagated splits are called the set of *available split* at the corresponding nodes in the partition tree. For each internal node of the partition tree, the split at the node is restricted to the subspace denoted by the node. To generate a new partition from a given one, one of its leaves (i.e. partition blocks) is divided into two new blocks using a split available at that node and the remaining splits from this set get propagated as described above. Starting with the single node partition tree which denotes the undivided space and applying a sequence of one or more node splits to this tree generates a new partition of the hierarchical space.

**Timestamped Partition Trees:** We associate a logical *timestamp* with each internal node of the partition tree. The timestamp represents the order in which the internal nodes were split. The tree shown in figure 3 is a timestamped partition tree, where the root has the timestamp of 1 (shown as a label in square brackets), the left child of the root node has a timestamp 2, the right child has a timestamp of 3, and so on. This implies that in creating the partition tree, the root was split first, followed by the left child, then the right child, etc. As will become clear, the concept of timestamp is important in preventing duplicate enumeration of the same hierarchical partition. Henceforth, by partition tree, we will refer to trees in which internal nodes are labelled by timestamp that represent the order in which the node was split to create the tree.

Now we describe our algorithm for the systematic enumeration of the hierarchical partitions that can be represented uniquely as the leaves of a multi-way "partition enumeration tree".

## 2.2 Partition Enumeration Tree (PET)

We first give the definition of a partition enumeration tree and then discuss how to generate it.

**Definition** 6. **(Partition Enumeration Tree):** *The partition enumeration tree (PET) is the multi-way tree, in which each node (including the root and leaves) denotes a distinct hierarchical partition of the space. Each node $n$ of PET corresponds to a unique timestamped partition tree denoted $P(n)$.*

Our objective is to generate a PET such that any two distinct nodes of the PET correspond to distinct space partitions. Nodes in PET are generated recursively by splitting one of the subspaces of the partition tree, $P(n)$, associated with a node $n$ of the PET using one of the split values that are available to split the subspace. We can avoid generating duplicate partition trees that are structurally similar to each other by exploiting the timestamps associated with nodes of the partition tree. This is achieved by imposing the following constraint on the partition trees associated with any node in the PET.

**Constraint** 1. *A partition tree associated with any node of a PET satisfies the*

*constraint that a* **pre-order** *traversal of its nodes (i.e. the partition tree's nodes) lists their timestamps in the increasing order.*

The above constraint prevents PET from generating nodes whose associated partition tree are structurally identical. However, as we observed in the example in the beginning of the section, the same hierarchical partitioning could result from two partition trees that are not structurally identical. Notice that the above simple constraint on timestamps would not prevent generation of this duplicate partitioning scheme. To prevent such duplicate generation of hierarchical partitioning, we need to define a concept of a in-sequence splits in a partitioning tree:

**Definition** 7. **(In-Sequence splits):** *Let $P$ be a partition tree, $n$ be a node of the tree, and $A$ be the set of ancestors of $n$ in $P$. The split $s_n$ associated with $n$ is* out of sequence *if there exists an ancestor $a \in A$ of $n$ such that: $s_n$ generates a $s_n$-cut across the subspace rooted at $a$ and the split $s_a$ associated with $a$ has a lower id than $s_n$. If a split $s_n$ is not out of sequence, it is deemed* in-sequence *for the partition tree.*

Duplicate generation of hierarchical partitioning in the PET can be prevented by imposing the following additional constraint on the partition tree $P(n)$ that correspond to the nodes of the PET.

**Constraint** 2. *A partition tree associated with nodes of a PET satisfies the constraint that the splits $s_k$ corresponding to any of its internal node $k$ are* in-sequence.

Consider the example of figure 3 again. Notice that the partition tree in the figure, seen without the node with timestamp [5] is legal. But, the final split (2), made at timestamp [5] would have been out of sequence since it results in a cut of the subspace rooted at its parent node (i.e., split(4)) while its id is smaller than the id of its parent.

## 3.  SEARCH FOR OPTIMAL PARTITIONING SCHEMES

The anonymization problem can be viewed as a cost based optimization problem. Without loss of generality we will assume that the optimization is a constrained minimization problem. Abstractly, an optimization problem is specified using two entities, *Cost* and *Set of Constraints*.

**Definition** 8. *Cost : $N \to \mathbb{R}$, where $N$ is the set of all nodes in PET, is a real function that associates each partition with a non-negative cost.*

**Definition** 9. Constraints: *A set of properties which determines the set of all feasible solutions. That is all potential solutions should minimally satisfy these constraints. We will also refer to an element from the feasible set as a* legal *solution.*

Different instances of these two parameters define different optimization problems. In context of the anonymization application, we define a few variants of the optimization problem using the following cost functions and constraints.
**Cost functions:** We consider the following cost functions:

—**Discernibility metric (DM):** Proposed in [Bayardo and Agrawal 2005], it assigns a penalty to each tuple based on how many tuples in the transformed data set are indistinguishable from it. If a tuple belongs to an anonymity group $S_i$ of size n (i.e. has $n-1$ other tuples present in it), then that tuple is assigned a penalty of $n$. If a tuple is suppressed a penalty equal to the size of the data set $| T |$ is assigned.

$$DM = \sum_{\forall i,\ |S_i| \geq k} | S_i |^2 + \sum_{\forall\ \text{Suppressed Tuples}} | T |$$

—**Classification metric (CM):** Also proposed in [Bayardo and Agrawal 2005; Iyengar 2002; Aggarwal and Yu 2004], captures the notion of information loss for classification based mining tasks. Assume $|T| = N$, then the information loss is given by:

$CM = \sum_{i=1}^{N} penalty(T(i))$ where $T(i)$ is the $i^{th}$ tuple and $penalty(t) = 1$ if $class(t)$ is not the majority class in the anonymity group of tuple $t$.

—**Volume metric (VM):** We propose a new metric to capture the notion of "relative increase in uncertainty" associated with the representation of a data point in the anonymized data set. For each tuple in a given equivalence class, a penalty $V/unit\_volume$ is charged, where $V$ is the "normalized volume" of the partition block to which the point belongs and $unit\_volume$ is the normalized volume of a basic unit cell of the data space (as a results of the initial discretization along the dimensions of the space).

$$VM = \sum_{i=1}^{N} Volume(block(T(i)))/emunit\_volume$$

where $Volume$ is the normalized volume function and $block(t)$ denotes the partitions block to which tuple $t$ belongs.

—**Count query estimate**: Consider a count query which results consist of the count of all tuples selected by the query predicates. To evaluate the utility of sanitized dataset we propose an empirical method that estimate the number of elements selected performing a range query $R$ on the anonymized data set, assuming that entries are uniformly distributed in each partition $P$.

$$Count\_Estimate = \sum_{\forall P_i \subseteq R} count(P_i) + \sum_{\forall P_i = leaf, P_i \cap R \neq \varnothing} density(P_i) \cdot overlap(P_i, R)$$

where $overlap(P, R)$ represents the volume intersection between the partition $P_i$ and the query region $R$ and $density(P_i) = count(P_i)/volume(P_i)$, is the number of tuples in each partition cell.

**Constraints:** Privacy definitions represent constraint settings which can be relevant to privacy preserving in data publishing. Our experiments are using different models, from the more popular, such as k-anonymity, l-diversity [LeFevre et al. 2006] up to the more recent ones [Li et al. 2007; Machanavajjhala et al. 2009].

—**k-Anonymity:** For every tuple $t'$ in the released table, there should at least be $l$ distinct indices (including its own) $(i_1, i_2, \ldots i_l)$, where $l \geq k$, such that $t_{i_1}(Q) = t_{i_2}(Q) = \ldots t_{i_l}(Q) = t'$, where $t(Q)$ denotes the projection of tuple $t$ on the quasi-identifier attribute set $Q$.

—**k-Anonymity with length restrictions:** Here in addition to the $k$-anonymity constraint we impose a minimum length constraint on the edge length(s) of a partition block in the final anonymized data set. Such constraints could reflect some security policy, wherein some of the quasi-identifying attributes are not allowed to be specified beyond a certain level of precision.

—**Entropy l-diversity:** Using the definition from [Machanavajjhala et al. 2007], a table is said to be entropy $l$-diverse if for every equivalence class (partition block) $b$, the following holds.

$$-\sum_{r \in R} p_{(b,r)} log(p_{(b,r)}) \geq log(l)$$

where $p_{(b,r)}$ denotes the fraction of tuples in the block $b$ with sensitive attribute value equal to $r$, and $R$ is the set of values (categorical) for the (single) sensitive attribute. This constraint ensures that each equivalence class has at least $l$ well represented values of the sensitive class.

—**t-closeness:** A microdata $T$ shows a particular distribution of the set $S$ of sensitive values $P(p_1, ... p_m)$, where

$$p_i = \sum_{s_i \in S} s_i / |T|;$$

each equivalence block $B$ shows this kind of distribution $Q(q_i, ..., p_m)$ as well, such as

$$q_i = \sum_{s_i \in S_B} s_i / |B|;$$

according to [Li et al. 2007] a table satisfies $t$-closeness if for each equivalence class, the distance between $P$ and $Q$ is no more than a threshold $t$. Such distance can be computed using Earth Mover's Distance (EMD) in such a way:

$$D[P,Q] = \frac{1}{m-1} \sum_{i=1}^{i=m} |\sum_{j=1}^{j=i} r_j|$$

where $m$ is the cardinality of $S$ and $r_i = p_i - q_i$.

—$\epsilon$ **-privacy:**    According to differential privacy definition, given two released tables $T_{in}$ and $T_{out}$, which contain and do not contain data about an individual respectively, the information gap between $T_{in}$ and $T_{out}$ should not be too much. As proposed in [Machanavajjhala et al. 2009], to guarantee $\epsilon$-privacy adversary's belief about victim's sensitive information $p^{in}$ and $p^{out}$, w.r.t. $T_{in}$ and $T_{out}$, has to be

$$\frac{p^{in}}{p^{out}} \leq \epsilon \qquad \text{and} \qquad \frac{1 - p^{out}}{1 - p^{in}} \leq \epsilon$$

According to the background knowledge that they have, adversary can be classified in four different classes, the first two called *realistic* as the amount of knowledge they have is finite whilst the latter two are called *unrealistic*, for the opposite reason. We consider the proposed constraint for *reaslistic* CLASS II adversary suitable for generalization techniques.

$$R1 \qquad n(q) - b \geq \frac{\sigma + b}{\epsilon - 1}$$

$$R2 \qquad \frac{n(q,s)}{n(q) + b} \leq 1 - \frac{1}{\epsilon' + \delta(q)}$$

$$where, \sigma(q) = (\epsilon - 1) \cdot \frac{n(q) - b}{\sigma + b} \quad \text{and} \quad \epsilon' = \epsilon \cdot \left(1 - \frac{1}{\sigma + b}\right)$$

A released table is $\epsilon$-private against a realistic adversary of CLASS II if for each equivalence class $q$ and for each sensitive value $s$ both condition R1 and R2 hold. In the above formula $n(q)$ is the cardinality of $q$, $n(q,s)$ is the distribution of $s$ in $q$, $\sigma$ is the number of total tuples learnt by the adversary from external sources (it's his *prior*).

**Definition** 10. **(Optimal Anonymization Problem):** *Given the above set of cost functions and constraints, for any combination of the two, find a hierarchical partition of the multidimensional data set such that the constraint is not violated for any partition block (equivalence class) and the cost of the solution is minimized.*

Now, we will describe how the above optimization problem can be posed as a search problem and illustrate our branch and bound technique to find an optimum solution.

### 3.1 Searching for the Optimum Solution

Given an instance of the optimization problem, we need to look for an optimal solution. A simple solution would be simply called algorithm **Enumerate** (figure 8) and return the least cost legal partition it generates[6]. The problem is that the search space for hierarchical partitions grows exponentially with the number of dimensions and number of splits, thereby making such a complete search impractical. More over not all enumerated partitions are necessarily legal, therefore to make the search tractable we employ cost lower bounding based node pruning. We also show how a nice structural property exhibited by our PET can be exploited to derive tight lower bounds.

It is clear that efficiency of the search algorithm critically depends on how much of the search space can be pruned away during enumeration. The following condition must hold for pruning a node $n$ (at its generation time) in our PET:

**Condition** 1. **(Condition for pruning)** *The subtree rooted at $n$ in the PET can be pruned if at least one of the following two conditions hold at each node in the subtree: (1) The partition corresponding to the node is not a feasible solution*

---

[6]This requires an additional constraint satisfaction test during generating a new partition in the Enumerate algorithm

*(violates one or more constraints); (2) The partition has a cost larger than the current_minimum_cost in the algorithm.*

The following monotonicity property of the constraints is essential for effective pruning of nodes in an enumeration tree based branch and bound approach.

**Property** 1. **(Monotonicity of a Constraint)** *Given optimization problem P, a constraint C and enumeration tree ET such that nodes ∈ ET correspond to elements of the solution space of P, then C is said to be monotonic with respect to ET if whenever a node $n \in ET$ violates C, all its descendants in ET also violate C.*

If a node $n$ in the search tree cannot be pruned for constraint violation, then we check for the second condition of prunability for the subtree rooted at $n$. To carry out this check an efficient (i.e., one that does not traverse the whole subtree) function $LB$ is required to estimate a lower bound to the minimum cost in the subtree. In our case, the function $LB$ is defined as follows.

**Definition** 11. *$LB : N \rightarrow \mathbb{R}$, where $N$ is the set of nodes in PET, is a real function that estimates a lower bound to the minimum cost for a hierarchical partition "derivable" from the partition at n (i.e. all partitions in the subtree rooted at n in PET).*

The $LB$ computation for a node can greatly benefit if the cost functions demonstrate the following monotonicity property with respect to the enumeration tree.

**Property** 2. **(Monotonicity of Cost)** *The value of the cost metric decreases monotonically down any path from the root to a leaf in the PET.*

Notice, that the monotonicity property of cost functions is a desirable property but not a necessity for computing lower bounds. For instance, in [Kifer et al. 2003] it is shown that in set mining problems, cost function like *variance* of values associated with items in a set do not display such monotonic behavior. The authors in [Kifer et al. 2003] go on to show some efficient ways of find a lower bound in such cases, but we will not consider such cost functions in this paper. All the cost functions and constraints that we mention above in the data publishing context, exhibit the monotonicity properties 2 and 1. The measures DM and CM have been shown to follow monotonicity for single dimensional partitions and the results can be easily extended to the multidimensional case. The new volume metric that we propose also exhibits a similar property due of the additive relation between the volume of a space and that of its constituent subspaces. Similarly, the privacy constraints, i.e., $k$-anonymity and $l$-diversity have been previously shown to be monotonic in [Bayardo and Agrawal 2005] and [Machanavajjhala et al. 2007] respectively. It is easy to see that this property extends to our new constraint of "$k$-anonymity with length constraints" as well.

In the following subsection, we describe the lower bounding methodology in the PET. We use the DM cost measure (defined earlier) for illustrating the approach.

3.1.1 *Cost based pruning in PET.* In addition to the monotonicity of our PET has a nice structural property that allows one to compute good lower bounds for all the cost functions described above. This property is potentially beneficial for lower bounding cost functions belonging to a rather well studied class of functions,

**LB**$_{DM}$(**m**)
**Input:** $m$ is some node in a partition tree
**Output:** Lower bound to the cost of any
partition of the space denoted by $m$
**BEGIN**
1. **If** $m$ is a leaf **Then**
2.     **Return** $MinCost(m)$
3. **End If**

4.     $m_r \leftarrow$ right-most internal node $\in subtree(m)$
5.     $L = \sum(Size(leaf))^2$, such that
    $leaf \in subtree(m)$ & $leaf \notin subtree(m_r)$
6.     $L \mathrel{+}= MinCost(m_r.rightchild)$
7.     **If** $m_r.leftchild$ is a leaf
8.         **Return** $L + MinCost(m_r.leftchild)$
9.     **Else**
10.        **Return** $L + $ **LB**$_{DM}(m_r.leftchild)$
11. **End If**
**END**

/* A node $m$ in a partition tree $P(n)$ denotes
a subspace. The subtree of $P(n)$ rooted at $m$,
referred to as $subtree(m)$, denotes the
restriction of the partition to this subspace.
$Size(leaf)$ is the number data points belonging
to the partition block denoted by $leaf$ */

Fig. 4.  Computing lower bound of $DM$

which happen to be some *spatial aggregates* over the data distribution across the partition blocks [Muthukrishnan et al. 1997]. We describe this property next.

**Property** 3. **(Spatial locality property)**: *Constraint 1 imposes restrictions on the set of subspaces that can be further partitioned at any descendant of a node in the PET. Specifically, consider any ancestor-descendant pair $n_a$ and $n_d$ in the PET and their corresponding partition trees $P(n_a)$ and $P(n_d)$, then $P(n_d)$ can differ from $P(n_a)$ only in the subtree rooted at the rightmost, non-leaf node $m_r$ of $P(n_a)$. That is, $m_r$ denotes the last internal node in $P(n_a)$ on the path starting at its root (inclusive) and containing only right branches of nodes, till a leaf is reached.*

We now illustrate the lower bound computation for DM metric (the others can be computed similarly) using the example given below and present the corresponding algorithm LB$_{DM}$ in figure 4.

The lower bound of the cost metric for a leaf $l$ of a partition tree is computed by the function call $MinCost(l)$ (illustrated below for the DM metric). It first generates the finest partition of the subspace denoted by $l$ by using all the available splits at $l$ (i.e. partitions the subspace into the smallest blocks possible) and then computes the value as follows:

$$MinCost(l) = \sum_{tuple \in subspace(l)} Penalty(tuple)$$

$$Penalty(t) = \begin{cases} \mid E(t) \mid & \text{where } \mid E(t) \mid > k \\ k & \text{otherwise} \end{cases}$$

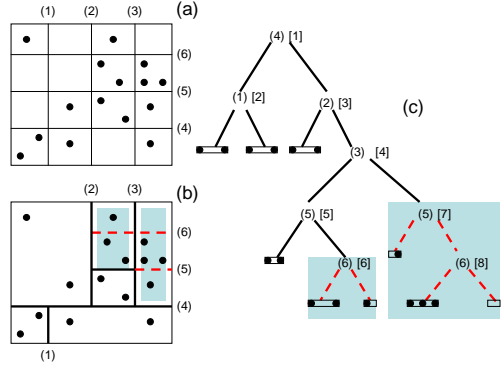

Fig. 5.   Visualizing lower bound computation

Where $\mid E(t) \mid$ represents the size of the partition block to which the data point $t$ belongs. That is, for each block having $m \geq k$ points, add $m^2$. For each block having $m' < k$ points, add $km'$. (Since the second set of blocks have less than $k$ elements, the anonymity criteria would have required one to suppress these elements. This would lead to a much larger increase of the $DM$ value, but since we are computing lower bounds, we simply charge each point a penalty of $k$, which is the minimum possible). We illustrate the lower bound computation using the example below (refer to figure 5).

**Example:** Consider the data space shown in the figure 5(a). Assume a node $n$ in PET corresponds to the partition shown in figure 5(b) without the "dashed" splits in the shaded region. Its partition tree is shown in figure 5(c) (visualized by replacing the 2 subtrees in the shaded region by 2 leaves instead). The splits used until this point are shown in bold, solid, black lines both in (b) and (c). The only legal partitions that our algorithm generates at any descendant of node $n$, will be by adding splits in the region shaded in blue in order to comply with the time-stamp constraint. The DM value of the partition $P(n)$ is given by: DM(n) = $2^2 + 2^2 + 2^2 + 2^2 + 3^2 + 4^2 = 41$ (leaves traversed in depth-first order). Now, the lower bound at $n$ is computed by first after adding all the available splits (generates 5 new leaves as shown in the shaded region) and summing the contribution from all the leaves of this partition: LB$_{DM}(n) = 2^2 + 2^2 + 2^2 + 2^2 + 2^2 + 1*2 + 1*2 + 3^2 + 0 = 33$. ◇

The lower bound computation for the other cost functions are similar to that of the DM metric.

## 4.   ENUMERATION ALGORITHM

Now we present our algorithm to enumerate duplicate-free hierarchical partitioning. The basic approach ensures that the partition tree associated with the nodes of the PET satisfy Constraints 1 and 2. The algorithm generates nodes in the PET recursively. Let $n$ be the current node of the PET being explored and let $P(n)$ be its corresponding partition tree. The partition tree associated with the child node of $n$ consists of $P(n)$ augmented with an additional split of one of the subspaces in $P(n)$. Generating such partition trees for child nodes to satisfy Constraint 1 (i.e., timestamp ordering) is straightforward. The constraint can be verified at the time the child node is created.

Checking if the partition tree associated with the child node satisfies Constraint 2 is more involved. Since it requires "stitching" splits across subspaces to detect out-of-sequence splits. We describe this next.

When a candidate split $s \in S$ is being considered at a leaf node in some partition tree $P(n)$, it is possible that a set of $s$-cuts across adjacent subspaces of the current block (denoted by the leaf) taken together form a $s$-cut of a larger subspace. The following two-part check is carried out by the function **Detect-legal-split** (figure 6: The function first calls the routine **Is-split-a-cut**$(P(n), t, s)$ (figure 7) which detects if the candidate split $s$ is a cut across a subspace denoted by node $t$ in partition tree $P(n)$. If the candidate split is indeed a cut of the subspace denoted by node $t$, its id is compared to the current node split at $t$ and a violation is returned if $s < split(t)$ (i.e., $s$ is out-of-sequence with $t$) else the same check is iteratively carried out at the parent of $t$ and so on till either a violation is detected or $t$ has no parent, i.e. node $t$ is the root of $P(n)$. The following observation is the key to an efficient implementation of this check which simply uses the information in $P(n)$ at a node $n \in$ PET.

**Observation** 1. *The partition tree $P(a)$ at an ancestor $a$ of node $n$ in PET is simply a tree generated by deleting a proper subtree of $P(n)$. Therefore $P(n)$ encodes all the information of its ancestors (nodes along the root-to-node $n$ path) in PET.*

The recursive enumeration algorithm **Enumerate** is given in figure 8. The algorithm is given the parameters: data set $D$, the ordered set of splits $S$, a node $r$ of the PET and the partition tree $P(r)$ at node $r$. When **Enumerate** is invoked with $r$ denoting the root of the PET, it generates the complete enumeration tree.

An example of a (partial) partition enumeration tree (PET) is shown in figure 9 for a data space with two attributes, one having 2 splits (1) and (2) and the other having a single split (3).

We can now state the main theorem that proves the correctness of our algorithm.

THEOREM 1. *Algorithm **Enumerate(D,S)** systematically enumerates all distinct hierarchical partitions of the multidimensional space $D$ generated using splits from the given split set $S$.*

Proof of the above theorem is given in the electronic appendix.

### 4.1 Categorical Attributes

In the description of the enumeration algorithms above, we have implicitly assumed that the attributes have an order defined on their domains. But in reality many attributes of a table can be categorical with absolutely no order defined on the values in its domain. Alternatively, there could be some partial order defined in the form of a lattice structure. A popular way of specifying a partial order over categorical attributes, is via a taxonomy tree. An example of a taxonomy tree is shown in figure 10 for the attribute *working class* in the "Adult" data set [KDD ]. The set of node labels of the taxonomy tree specifies the domain for the corresponding categorical

---

**Detect-legal-split($\mathbf{P(n)}$, $\mathbf{l}$, $\mathbf{s_l}$)**
**Input:** $P(n)$ is partition tree of node $n \in PET$
           $l$ is the leaf being split in $P(n)$
           (i.e. $l$ denotes an unsplit partition block)
           $s_l$ is a candidate split at $l$
**Output:** True(False) if split $s_l$ is Legal(Illegal)

**BEGIN**
1.    **If** $l == root(P(n))$ **Then**
2.      **Return** TRUE
3.    **End if**

4.    **If** Pre-order listing of node time-stamps in
5.      $(P(n) \cup s_l)$ NOT in increasing order **Then**
6.        **Return** FALSE
7.    **End if**

8.    $a \leftarrow l$
9.    **While** $a \neq root(P(n))$ **Do**
10.     $a \leftarrow parent(a)$
11.     **If** Is-split-a-cut$(P(n), a, s_l)$ **Then**
12.      **If** $s_l \leq split(a)$ **Then**
13.        **Return** FALSE
14.      **End if**
15.     **Else** /* $s_l$ is not a cut of subspace of $a$ */
16.      **Return** TRUE
17.     **End if**
18.    **End While** /* now $a = root(P(n))$ */
19.    **Return** TRUE
**END**

---

Fig. 6.    Detecting a legal split

attribute. For instance, the *working class* attribute of a tuple can take values only from the set of node labels in the taxonomy tree.

Our enumeration algorithm can handle categorical attributes easily with a couple of slight modifications to the approach outlined for numeric attributes. We categorize the categorical attributes into the following two classes and specify the corresponding modifications that are required to split along these dimensions of the space.

(1) **No order:** In this case, no order is defined on the set of values taken by the attribute. As a result, all possible groupings are possible. For example, say the attribute takes 3 distinct values $a$, $b$ and $c$, therefore blocks (anonymity groups) in a partition can cluster together these values in any of the following 7 $(= 2^3 - 1)$ ways: $\{(a,b,c);(a,b);(a,c);(b,c);(a);(b);(c)\}$, where all values within '(' and ')' are indistinguishable. This case is equivalent to having three independent binary attributes instead of one with three values.

(2) **Partial order:** If the categorical attribute is denoted by $C$, in this case we assume a taxonomy tree $T_C$ is defined for the values that $C$ can take. The following additional constraints need to be imposed while splitting a set of data points along the attribute $C$ (refer to figure 10): (i) For a given generalization scheme, induced by some partition, all tuples in a partition block (anonymity group) share the same values of the quasi-identifier attributes. Therefore, attribute $C$ for each tuple in a block $l$ should correspond exactly to one node $t_c$

**Is-split-a-cut(P(n), t, s)**
**Input:** $P(n)$ is partition tree of node $n \in PET$
         $t$ is an internal node of $P(n)$ (can be the root)
         $s$ is some splitting attribute value
**Output:** True if there is a $s$-cut across the
          subspace denoted by $t$ else False

**BEGIN**
1.     **If** $split(t) \in group(s)$ **Then**
/* Here, we need a small test to determine
if some split $s'$ is in the same group as
split $s$ (i.e., belongs to same attribute) */

2.        **If** $split(t) == s$ **Then**
3.           **Return** TRUE
4.        **Else If** $split(t) < s$ **Then**
5.           **Return** Is-split-a-cut($P(n), rightchild(t), s$)
6.        **Else** /* $split(t) > s$ */
7.           **Return** Is-split-a-cut($P(n), leftchild(t), s$)
8.        **End if**
9.     **Else** /* $split(t) \notin group(s)$ */
10.       **If** $leftchild(t)$ is a leaf **OR**
          $rightchild(t)$ is a leaf **Then**
12.          **Return** FALSE
13.       **Else**
14.          **Return** (Is-split-a-cut($P(n), leftchild(t), s$)
15.             **&** Is-split-a-cut($P(n), rightchild(t), s$))
16.       **End if**
17.    **End if**
**END**

Fig. 7.   Detecting if a new split results in a cut across a larger enclosing sub-space

**Enumerate(D, S, r, P(r))**
**Output:** The enumeration tree rooted at $r$,
          consisting of all distinct partitions legally
          generatable from the partition $P(r)$

/* Below, $s_l$ ranges over only the "available"
splits at the leaf $l$ (i.e. only the splits relevant
to the partition-block denoted by $l$). */

**BEGIN**
1.  **For Each** leaf $l$ of $P(r)$ /* in pre-order sequence */
2.     **For Each** available split $s_l$ at $l$ /* increasing order of id */
3.        **If Detect-legal-split**($P(r), l, s_l$) **Then**
4.           Generate new child node $c$
5.           Generate partition-tree $P(c) \leftarrow P(r) \cup s_l$
6.           Add pointers $r \rightarrow c$ ; $c \rightarrow r$
7.           **Enumerate**($D, S, c, P(c)$)
8.        **End if**
9.     **End For**
10. **End For**
**END**

Fig. 8.   Enumerating all partitions of space

of the taxonomy tree $T_C$ (say "Government"). Let the set of these tuples be
$D_l$. (ii) Subsequent splitting of the block $l$ along attribute $C$ should **simultaneously specialize** the $C$-attribute values of tuples in $D_l$ to the children of node $t_c$ in $T_C$. For example, splitting $D_l$ would result in groups $D_{l1}, D_{l2}$ and $D_{l3}$ with *working class* values as "Fed", "State" and "Local" respectively, where $S = D_{l1} \cup D_{l2} \cup D_{l3}$ and $D_{li} \cap D_{lj} = \phi, \forall i \neq j$.

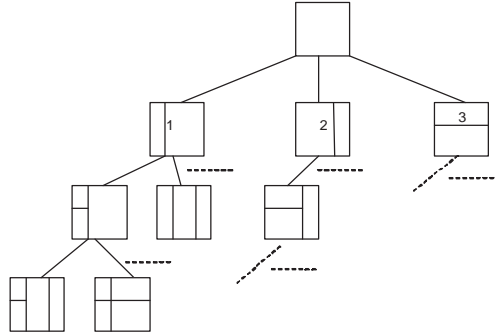

Fig. 9.    Partition enumeration tree

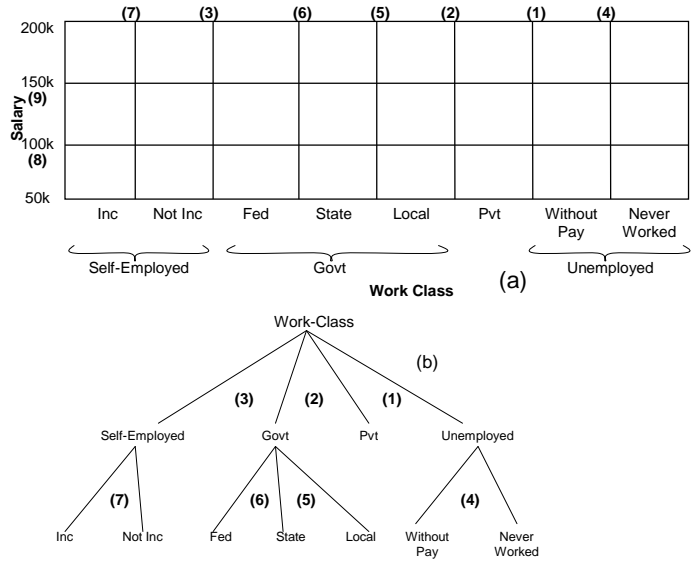

Fig. 10.    (a) Split priorities for a categorical attribute; (b) Priorities for taxonomy-tree nodes

Tackling categorical attributes without any order whatsoever is easy since such an attribute can be replaced by a set of binary attributes. In this paper, we assume each categorical attribute has a partial order defined on them, in the form of a taxonomy tree. Recall that in the enumeration algorithm, new partitions are generated from a given partition by splitting some partition block of the parent partition into two blocks by introducing exactly a single split along some chosen dimension. This allows us to represent each new partition also as a binary tree. But introducing the splitting constraint ((ii) in item 2 above) can result in more than two blocks being generated simultaneously when a partition block of the parent partition tree is split along a categorical attribute. Say, if the node corresponding to the current value of the categorical attribute $C$ has $m$ children in its taxonomy tree $T_C$, splitting along this attribute will result in $m$ leaves simultaneously in the new partition tree. Therefore to represent this new partition in the standard format of our partition tree, we need to assign priorities to the $m$ children of the corresponding node in the taxonomy tree. Two considerations need to be made while introducing these splits into the parent partition tree: (i) The "child priority" constraint 1 and "Out-of-sequence split" constraint 2 ordering constraints are not violated in the resulting partition tree due to introduction of these sibling splits (i.e. sibling nodes in $T_C$). (ii) Future splits in the newly generated blocks should not be wrongly discarded (i.e., for violating either the time-stamp constraint or the split priority constraint). These two requirements needs one to assign: (a) Consecutive split priorities to each set of sibling splits. (b) Also, assuming the "**left-to-right**" listing of the sibling nodes in $T_C$ correspond to the actual left to right ordering of the splits in the space[7], the assigned priority should be **decreasing** amongst sibling splits going from **right to left** (as shown in figure 10); (c) The sibling splits (in $T_C$) need to be introduced simultaneously as noted above, but we also need to retain the binary tree form of the partition tree. As a result the sibling splits need to be introduced in their decreasing order of priority (i.e., high priority splits first). This results in creating a **left deep** binary subtree under the leaf being split and therefore complies with the time-stamp constraint. Now, with the following example, we describe the splitting technique. A more accurate description of the algorithm in Fig 12 is given in the subsequent paragraph.

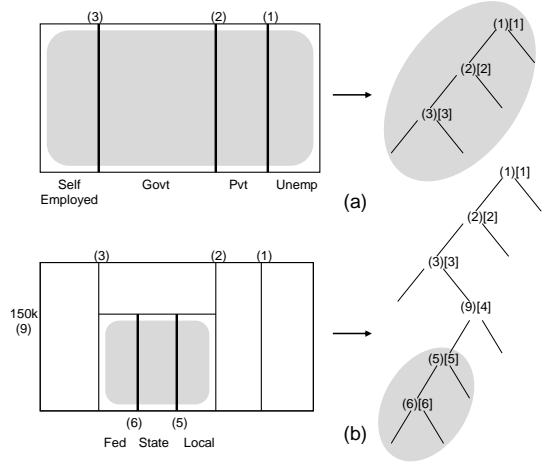

Fig. 11. Splitting along categorical attributes can introduce multiple nodes in partition tree in one step

**Example:** Assume we have a two dimensional data set with one dimension as numeric attribute *salary* with four values (intervals) and the other, the categorical

---

[7]remember that each attribute has a well defined notion of "left end" and "right end" along the dimension corresponding to it in the data space

attribute *working class* with a taxonomy (figure 10, part (b)) defined on it. Part (a) of figure 10 shows the space and all the available split values with their priorities assigned according to the rules mentioned above. Initially assume that the space is undivided therefore all data elements are indistinguishable from each other. Now let the first split be introduced along attribute *working class* which results in creation of the four blocks in one go (figure 11, part (a)) thereby adding 3 nodes to the partition tree, in the left-deep manner (note this complies with the time-stamp ordering constraint as well as the out-of-sequence split constraint). In part (b) of figure 11, say a split along the *salary* attribute is chosen in the second block from left which adds node "(9)[4]" (split (9) and time-stamp [4]). Then, splitting the lower of the two newly generated blocks (shaded portion) along the categorical attribute, results in three more blocks in one go, i.e. 2 new nodes in the partition tree (time stamps [5] and [6]).

The algorithm in Fig 12 illustrates the specific steps required to perform a correct partitioning when categorical attributes are involved. It is important to note that when performing a split on a categorical dimension can require subsequent splits. For instance, in Fig 10 performing a split on (1) forces the splits (2) and (3). Further, note that it is not possible to have a partitioning using splits (4), (5), (6) and (7) without using the splits (1), (2) and (3) first. Therefore, the *Detect-legal-split* routine has to implement these extra constraints in order to incorporate categorical attributes.

The function *hasHierarchySplits(P(r))* indicates whether the current partition tree $P(r)$ needs more splits to complete the partitioning on the categorical attribute (e.g. in 11 split (1) has been performed but split (2) has not yet) or not. The function *needMoreSplits($s_l$)* indicates whether the splits $s_l$ has required right-most siblings: for instance split(1) has siblings (2) and (3), but split (3) has no siblings. Using such functions, the algorithm checks whether we need to keep splitting on a categorical attribute or not.

## 4.2 Accelerating Search using Priorities

As it turns out, more often than not in spite of pruning, the search space continues to remain extremely large. As a result, the depth-first traversal order of nodes (as in the recursive algorithm) turns out to be an inefficient way to explore the solution space for most objective functions of interest. Here, by efficiency we mean "how soon the algorithm finds a solution close to optimal", say a partition whose cost is within a small factor $\alpha$ of the minimum cost. To do away with these limitations of the recursive algorithm, we propose a new, generic solution-space exploration scheme that is both theoretically sound (i.e. guarantees completeness of search just as the recursive enumeration algorithm) and at the same time, allows one to incorporate heuristics that accelerate the convergence to the optimum solution, the side-effect being: good approximations to the optimal are generated much quicker. Now, instead of traversing the PET in a fixed order which is agnostic of the objective function, our new algorithm uses a flexible, *priority* based search scheme to direct its search at each step.

**Definition** 12. *Priority* : $N \rightarrow \mathbb{R}$, *where $N$ is the set of nodes in PET, is a real function which assigns a numeric value that is used as the key to insert $n$ into*

**Enumerate Categorical($\mathbf{D}$, $\mathbf{S}$, $\mathbf{r}$, $\mathbf{P(r)}$)**
**Output:** The enumeration tree rooted at $r$,
        consisting of all distinct partitions legally
        generatable from the partition $P(r)$
        using categorical attribute

/* Below, $hasHierarchySplits$ shows if $P(r)$ needs
more splits to complete the hierarchical partitioning;
$needMoreSplits(s_l)$ indicates if the cut $s_l$ has siblings */

**BEGIN**
1.    **For Each** leaf $l$ of $P(r)$ /* in pre-order sequence */
2.      **If**(hasHierarchySplits(P(r)) **Then**
3.        $availableSplits(l) \leftarrow getNextSplits(P(r).split)$
4.      **End if**
5.      **For Each** $availableSplit(l)$ $s_l$ /* increasing order of id */
6.        **If Detect-legal-split**($P(r), l, s_l$) **Then**
7.          **If**($s_l \in Hierarchy\ h$) **and** $needMoreSplits(P(r))$ **Then**
8.            hasHierarchySplits(P(r)) $\leftarrow$ TRUE
9.          **Else**
10.           hasHierarchySplits(P(r)) $\leftarrow$ FALSE
11.          **End if**
12.          Generate new child node $c$
13.          Generate partition-tree $P(c) \leftarrow P(r) \cup s_l$
14.          Add pointers $r \rightarrow c$ ; $c \rightarrow r$
15.          **Enumerate**($D, S, c, P(c)$)
16.        **End if**
17.      **End For**
18.    **End For**
**END**

Fig. 12.   Enumerating all partitions of space with categorical attributes

the priority queue. We also refer to it as the priority of node $n$.

Our algorithm admits the usage of arbitrary priority generating functions. We experimented used the following functions to generate node priorities (i.e., instances of the "$Priority$" function defined above).

**1)** $LB$ (lower bound function): $LB$ was the primary priority generating function and was used in majority of the experimental runs. The rationale being that, the branch of PET that has lowest lower bound is the most promising one to explore.

**2)** $Cost$ (cost function): Cost of the solution at a node $n$ was used to prioritize the search. The goal was to see how going down a path with the minimum cost affects the quality of solutions generated.

**3)** $LB/Cost$ (ratio of the lower bound and the cost): Similarly, the rationale behind $LB/Cost$ was to go down branches which had the highest potential of cost improvement.

The experimental results for these functions are summarized in the next section. Now, we describe the new search algorithm using the priority queue which was the algorithm used in all our experimental runs. There are two modes in which the priority algorithm can be run: (A) Geared towards finding **an optimum solution**, where the lower bound $LB(n)$ is used to decide whether to prune node $n$ or not. (B) Geared towards finding a $c$-**approximate solution** to the optimum, where the ratio $\alpha(n) = LB(n)/Cost(n)$ is used to prune node $n$ from the search space. For instance, if the user wants to find a 3-approximate solution, a node with $\alpha$

---

**Prioritized-Enumerate(D, root)**
**Input:** $D$ is the data set to be partitioned
      $r$ is the root node in $PET$
**Output:** Output the optimal partition and its cost
**BEGIN**
1.    $OPT_{tree} \leftarrow P(root)$
2.    $OPT_{value} \leftarrow Cost(P(root))$
3.    **If** $Opt_{value} < LB(root)$ **Then**
4.       **Return** $OPT_{value}$ and $OPT_{tree}$
5.    **End If**

6.    $PQ \leftarrow$ insert$(root, LB(root))$
7.    **While** $PQ \neq \phi$ **Do**
8.       **If** $|PQ| \geq$ MAXSIZE **Then**
9.          **While** $|PQ| >$ MAXSIZE/2
10.            $R_p \leftarrow$ A good partition beneath $min(PQ)$
11.            $OPT_{value} \leftarrow Cost(R_p)$ and $OPT_{tree} \leftarrow R_p$
12.            Delete all $n'$ in $PQ$ with $LB(n') > OPT_{value}$
13.          **End While** /* Now $|PQ| \leq MAXSIZE/2$ */
14.       **End If**

15.       **While** $|PQ| \leq MAXSIZE$ & $PQ \neq \phi$ **Do**
16.          $x \leftarrow PQ.pop()$
17.          $X_c \leftarrow x.children$ /* $x$ a single split */
18.          insert into $PQ$, all $y \in X_c$ s.t $LB(y) < OPT_{value}$
19.          **If** $Min_{y \in x \cup X_c} Cost(y) < OPT_{value}$ **Then**
20.            update $OPT_{value}$ and $OPT_{tree}$
21.          **End If**
22.       **End While**
23.    **End While**
24.    **Return** $OPT_{value}$ and $OPT_{tree}$
**END**

---

Fig. 13.   The prioritized search algorithm

value greater than 1/3 can be safely pruned since one would have already seen a candidate solution if indeed the true minimum was same or higher that the lower bound at the node (the current global minima being one such candidate). Let us use the variable *bound* to denote either *LB* or $\alpha$ depending on the mode of the algorithm's run, then the basic priority queue does the following.

**Priority-queue based search algorithm:** The algorithm starts with the root node of the PET as the singleton node in the queue and the value in the variable *current_minimum_cost* set to $Cost(root)$, which is the cost when the whole data set is a single partition block. In each successive step, the top element of the queue[8] is popped and all its children are generated. Some of these newly generated nodes using their priorities. For each new node, if the cost is found to be lower than the value in the variable *current_minimum_cost*, this value is replaced by the new minimum and the new optimal partition is recorded. Also, for each new node $n$ *bound(n)* depending on the mode in which it is being executed. If the *bound(n)* is higher than the current global minima, this node is discarded[9]. The algorithm terminates when the priority queue is empty and the partition corresponding to the current minima is the optimal solution. The pseudo-code for the priorities based algorithm is given in figure 13.

---

[8]We use a MIN-priority queue, where the root node always has the lowest value of the key amongst all nodes
[9]It is quite possible that the partition corresponding to a discarded node might be the best solution seen so far

**Practical issues in using a priority queue:** The downside of using a priority queue based approach is that the queue can grow very large in size on certain occasions. In such circumstances, the following probe-based algorithm is used.

**Probe-based algorithm:** When max-allowed size is reached, this algorithm does a "probe" which temporarily puts on hold further enumeration and instead invests time to find a node in the PET that has a cost smaller than the lower bound of a large number of nodes in the priority queue. Then, these nodes become useless and can be pruned from the queue. In the event such a good partition cannot be found by probing, the alternative is to forcefully drop a certain number of entries that are the least promising, thereby forfeiting the promise of optimality of the final solutions. Nonetheless, due to the inherent nature of the prioritized algorithm, a bound on the approximation ratio achieved by any candidate solution can always be derived. Note that the efficiency of the above algorithm depends upon the choice of parameter MAXSIZE and we will present some of our experimental findings regarding the performance with respect to MAXSIZE in the next section.

In the next section, we discuss the various experiments that we carried out.

## 5. EXPERIMENTS

**Experimental setup:** All the experiments were run on a Pentium 4, 3.00 GHZ processor machine with 4 GB RAM. The machine was running windows XP operating system and our enumeration algorithm was implemented using the VC++ development environment.

**Data sets:** There were two different datasets that were used in the experiments. Both datasets are from the Irvine machine learning repository [KDD ]. The first real data set used in the experiments is the *Adult* data set, which has in some ways become the benchmark for comparing anonymization algorithms. This gives us the opportunity to compare the quality of our solutions against other algorithms/techniques previously proposed. This data has 9 attributes (Age, Geography, Gender, Race, Working Class, Occupation, Education, Class, Marriage) and reports the actual census data. The data set has 30,162 tuples.

The second real data set used in our experiments is the *Census-Income* from the Irvine machine lerning repository. This data set contains weighted census data extracted from the 1994 and 1995 current population surveys conducted by the U.S. Census Bureau. The data set contains 199523 records. We have considered 4 *Quasy-Identifiers* attributes (*age*, *working class*, *marital status*, *gender*) and one *Sensitive* attribute (*income*).

**Purpose of the experiments:** The purpose of our experiments is to measure: a) scalability of our enumeration algorithm; and b) comparison of enumeration algorithm with other approaches in the literature. Primarily, we compare ourself with the greedy algorithm proposed in [LeFevre et al. 2006], as the this approach outperforms other previously suggested techniques. We will show that the gap between our approach and previous techniques performances in terms of information utility increases as privacy constraints become more stringent. Our approach can give good results even with very tight privacy requirements.

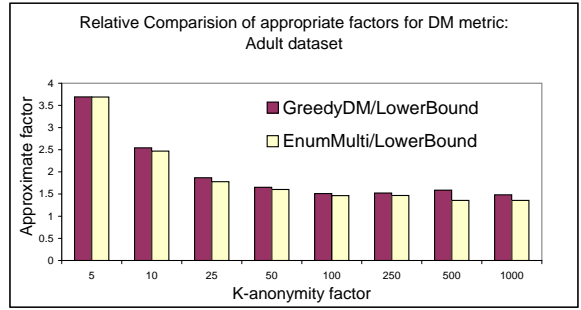

Fig. 14. Performances on Adult dataset and DM

**Scalability:** We empirically test the performance of our enumeration algorithm, specifically the time it takes to find the optimal (in most cases, we report how close it approximately gets the to the theoretical best-possible solution when it cannot find the optimal solution in a given maximum length of the run). The most critical parameter for the enumeration algorithm is the number of splits. The number of splits dictate the solution space. We have found that the enumeration algorithm finds optimal solution under 30 minutes when the number of distinct splits used are less than equal to 10. When the number of splits were increased, we needed to drop some solutions as they could not fit in the memory and hence could not guarantee the optimality of the final solution. When the splits were more than 10, the algorithm finds solutions within a factor of 3.5 of the theoretical lower bound to the cost. Lower bound is calculated by assigning the lowest penalty possible to every tuple. For instance, for the DM metric, the lower bound is $n * k$, where $n$ is the number of tuples and $k$ is the anonymity factor. Whereas our enumeration algorithm can determine the optimal solution or those within a specified factor of the lower-bound to the theoretical best given sufficient space and time, the greedy algorithms proposed thus far typically return a locally optimal solution.

**Comparison with greedy algorithms:** Comparison to the greedy is done for 3 different metrics (DM, VM, CM) and different constraints different constraints ($k$-anonymity, $l$-diversity, $t$-closeness and $\epsilon$-privacy (class II adversary)). Each of these constraints have a parameter that is varied, $k$ for anonymity, $l$ for diversity, $t$ for closeness and three parameters ($\epsilon$, $\sigma$ and $B$) for privacy against realistic adversaries. We applied different sets of constraints on each combination of dataset and cost metric. For lack of space, we will only report experiments that are done for the DM metric in this paper. We have also evaluated the quality of anonymized data in terms of estimate of count queries results on such sanytized data. For all the following experiments the enumeration algorithm was run for maximum of 5 hours.

**Experiment 1:- (K-anonymity on adult dataset, DM):**
We used 101 splits to partition the adult dataset. Our enumeration technique outperforms the greedy approach at higher values of $k$(greater than 100) and does marginally better at lower values of k. To capture how close we are getting to the optimum solution, we calculated the approximation factor which is the ratio of the best solution found to the lowest lower bound of a partition tree enumerated by our algorithm[10]. Similarly, we have calculated the approximation factor for the greedy approach and fig 14 shows the comparison of the approximate factors for the two approaches. We have got as close as 1.35 times the lowerbound (for K=1000).

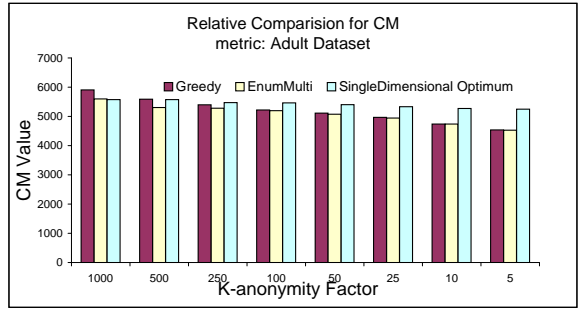

Fig. 15.   Performances on Adult dataset and CM

The most important result to come out of this experiment is the fact that greedy algorithm comes up with close to optimal solutions when the dataset is partitioned using a large number of splits. This confirms the intuition that we have gained from the previous experiment, where greedy algorithm was getting closer to the optimal solutions, when the number of splits increase. The greedy algorithm at every stage tries to find the best split to break a partition into two pieces. If there are a large number of splits at its disposal, even at lower levels the greedy algorithm will find a good candidate split. Since the DM metric is monotonically decreasing, as long as the greedy algorithm finds a split, the value of the DM metric value will decrease.

**Experiment 2:- (k-anonymity on Adult dataset, CM):**

For this experiment we have used 101 splits. The enumeration algorithm mimics the result of the DM metric for the CM metric. The enumeration algorithm does better than greedy for higher values of k, but the greedy algorithm comes close to the enumeration algorithm for lower values of k. Fig 15 shows the result of the experiment.

**Experiment 3:- (Combination of privacy constraints on Census-income dataset, DM):**

This experiment has been carried out on Census-income data set with an overall of 30 splits using different combination of privacy constraints: $\sigma$, $\epsilon$ and $b$ as seen in section 3 are all $\epsilon$-privacy parameters. Fig 16 shows that our algorithm works up to 3 times better than *Mondrian* with different combination of constraint parameters. Even when constraints are very stringent (first histogram) *GenSearch* works almost 50% better.

**Experiment 4:- (Combination of privacy constraints on Adult dataset, DM):**

This experiment refers to Adult data set with 8 Quasi-Identifier attributes, 1 sensitive attribute and 102 splits. Results, shown in Fig 17, are similar to the ones obtained in the previous experiment.

**Experiment 5:- (Different Split sets on Census-income dataset, DM):**

---

[10]This is also lower bound of the optimal solution

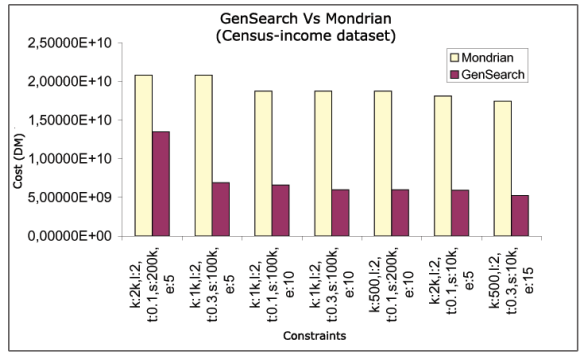

Fig. 16. Comparison of Mondrian and GenSearch algorithms using DM and various combination of privacy constraints on Census-Income dataset

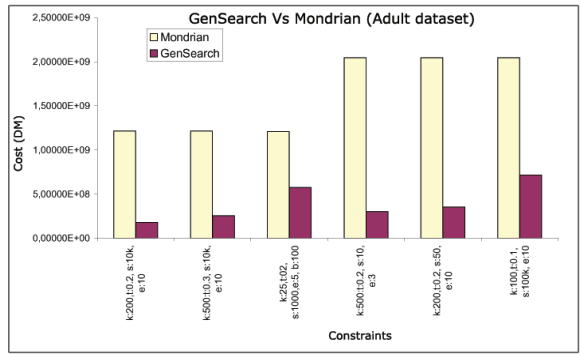

Fig. 17. Comparison of Mondrian and GenSearch algorithms using DM and various combination of privacy constraints on Adult dataset

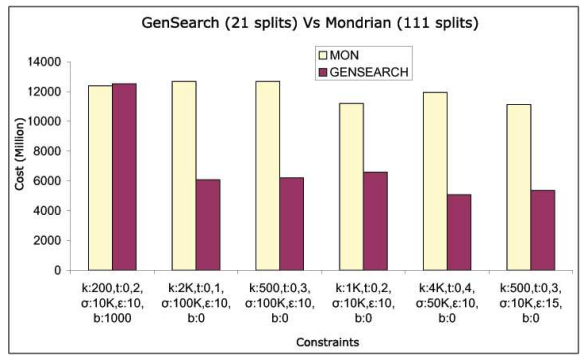

Fig. 18. Performance of GenSearch and Mondrian using different Split Sets on Census-Income Dataset

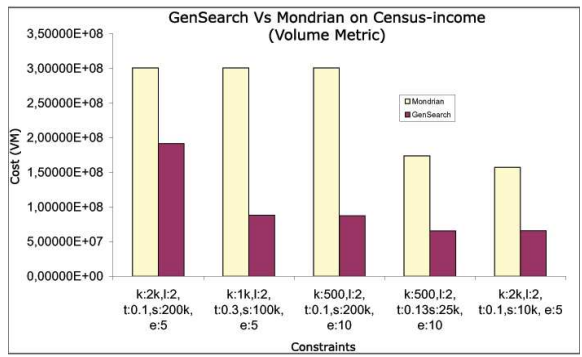

Fig. 19. Performance of GenSearch and Mondrian using Volume Metric on Census-Income Dataset

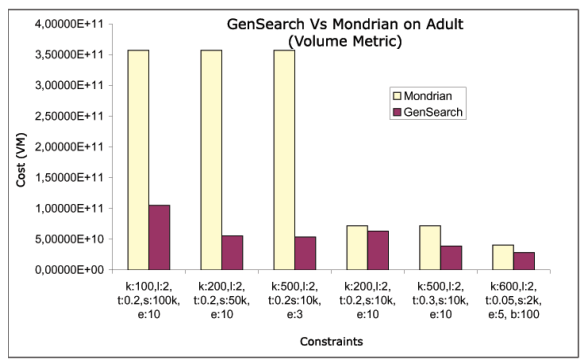

Fig. 20. Performance of GenSearch and Mondrian using Volume Metric on Adult Dataset

Using a large number of splits leads generally to a better solution, as the dataset discretization is more accurate. In Fig 18 we can see how *GenSearch* works better than previous multidimensional approach even using a smaller *Split Set* size. In such experiment we used 21 splits for *GenSearch* and 111 splits for *Mondrian*. The results shows that in one case *Mondrian* works better, but the gap from corresponding *GenSearch* results is negligible. However, *GenSearch* works an average of two times better than *Mondrian*.

**Experiment 6:- (Combination of privacy constraint on Census-income dataset, VM):**

Other experiments have been carried out on Census-income dataset considering Volume Metric as cost function. Results in Fig 19 shows that *GenSearch* works better than *Mondrian* for different combinations of privacy constraints, even with respect to *Volume Metric*. *GenSearch* outcomes have been obtained after executing such algorithm for one hour.

**Experiment 7:- (Combination of privacy constraint on Adult dataset, VM):**

*GenSearch* algorithm applied on Adult dataset, with a set of 66 splits, gives better results than *Mondrian* with respect to Volume Metric function cost, as shown in Fig 20. As in the previous experiment, we ran *GenSearch* for one hour

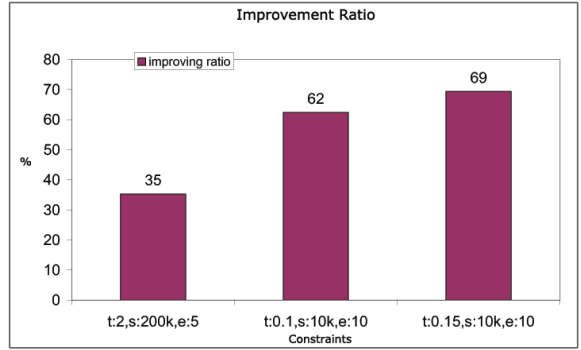

Fig. 21.   Improving ratio starting from Mondrian initial solution, Census-Income Dataset

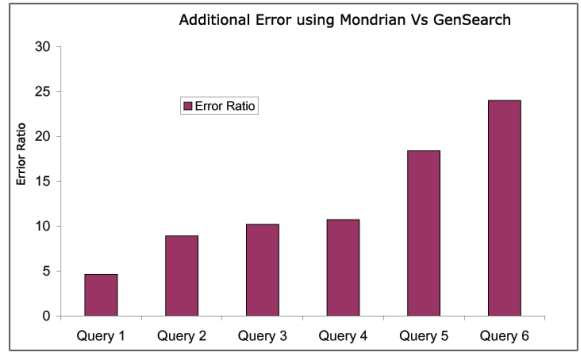

Fig. 22.   Error gap of count queries estimate between GenSearch and Mondrian approach

to obtain the shown outcomes values.

**Experiment 8:- (Improving ratio starting from Mondrian solution, DM):**

This experiment aims to evaluate the improvement of cost values got by *GenSearch* using *Mondrian* solution as starting point. *GenSearch* can improve solution value up to almost 70% as shown in Fig 21.

**Experiment 9:- (Estimate of count query results on sanitized Adult data set):**

This experiment has been carried out on Adult data set sanitized using $k = 500$, $l = 2$, $t = 0.2$, $\sigma = 10K$, $\epsilon = 10$ and $b = 0$ as privacy parameters for *GenSearch* and *Mondrian*. *GenSearch* released data set has been generated after 10 minutes. The experiment wants to evaluate the error in estimate the counts of records given by a SQL style selection predicate. In the following there are listed the queries which results are shown in Fig 22:

*Query 1*

```
SELECT COUNT(*)
FROM ADULT A
```

```
WHERE A.MARITAL IN ('MARRIED-CIV-SPOUSE', 'NEVER-MARRIED',
                    'MARRIED-SPOUSE-ABSENT', 'DIVORCED')
AND A.ETHNICITY IN ('WHITE', 'BLACK', 'ASIAN-PAC-ISLANDER',
                    'AMER-INDIAN-ESKIMO')
```

*Query 2*

```
SELECT COUNT(*)
FROM ADULT A
WHERE A.AGE BETWEEN 37 AND 77
AND A.MARITAL IN ('MARRIED-CIV-SPOUSE', 'NEVER-MARRIED',
                  'MARRIED-SPOUSE-ABSENT', 'DIVORCED')
AND A.ETHNICITY IN ('WHITE', 'BLACK', 'ASIAN-PAC-ISLANDER',
                    'AMER-INDIAN-ESKIMO')
```

*Query 3*

```
SELECT COUNT(*)
FROM ADULT A
WHERE A.AGE BETWEEN 22 AND 42
AND A.ETHNICITY IN ('WHITE', 'BLACK')
```

*Query 4*

```
SELECT COUNT(*)
FROM ADULT A
WHERE A.AGE BETWEEN 47 AND 67
AND A.MARITAL IN ('MARRIED-CIV-SPOUSE', 'NEVER-MARRIED',
                  'DIVORCED')
AND A.ETHNICITY IN ('WHITE', 'BLACK')
```

*Query 5*

```
SELECT COUNT(*)
FROM ADULT A
WHERE A.AGE BETWEEN 42 AND 52
AND A.MARITAL IN ('MARRIED-CIV-SPOUSE', 'DIVORCED',
                  'MARRIED-SPOUSE-ABSENT')
AND A.ETHNICITY IN ('WHITE', 'BLACK', 'ASIAN-PAC-ISLANDER')
```

*Query 6*

```
SELECT COUNT(*)
FROM ADULT A
WHERE A.AGE BETWEEN 42 AND 52
AND A.MARITAL IN ('MARRIED-CIV-SPOUSE', 'NEVER-MARRIED')
AND A.ETHNICITY IN ('WHITE', 'BLACK')
```

The experiment shows how worse the estimate of such queries is, using *Mondrian* approach instead of *GenSearch*.

**Miscellenaous experiments:**

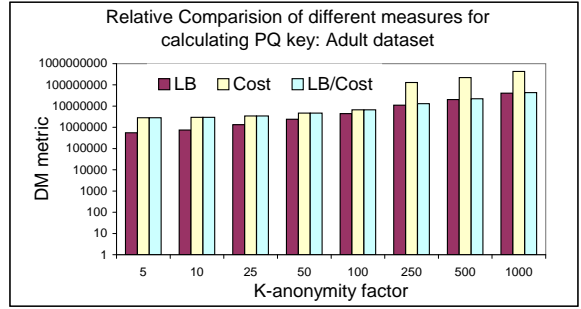

Fig. 23.   Biasing Priority queue

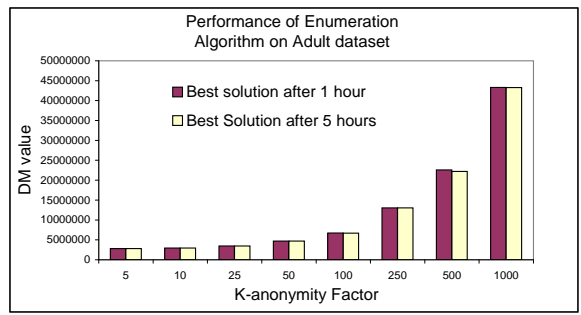

Fig. 24.   Performance, Adult Dataset

Another interesting measurable factor was the priority function(key) of the priority queue. Changing the key of the priority queue leads to interesting search patterns on the partition enumeration tree and could potentially increases the chances of finding good solutions quickly. Previously, we introduced the following three different priority functions a) LowerBound ($LB$) b) Cost of the metric ($Cost$) and c) $LB/Cost$ and stated their rationale. Fig 23 shows the result for a run on the adult dataset optimizing the DM metric using all the three priority functions(keys) under consideration. $LB$ priority function performs the best. A good solution at the highest level of the PET does not necessarily lead to a good solution at the lower levels. Therefore, $Cost$ priority function does not do as well the $LB$ priority function. In the case of $LB/Cost$ priority function, the Cost of the metric ($Cost$) dominates the value and the priority function follows nearly the same search pattern as the $Cost$ priority function.

One interesting fact about the enumeration algorithm is that it does not need to run for a long time to get good solutions. In Fig 24 we have plotted the best solution found after 1 hour and after 5 hours for a run on the adult dataset optimizing the DM metric. It is quite clear from Fig 24 that the metric value of the solution found after 1 hour is very close to the metric value of the solution found 5 hours. This characteristic is found in almost all of our experiments.

In summary, the following are the significant results of our experiments;

—The enumeration algorithm has found optimal solutions for up to 14 splits under

30 mins.

—The enumeration algorithm performs better than greedy in all experiments and performs significantly better than greedy for higher values of k (usually more than 100), when constraints are not applied.

—For lower values of k (less than 100), greedy comes close to solutions found by the enumeration algorithm, when constraints are not applied.

—When constraints are applied, enumeration algorithm outperforms the greedy techniques for all values of k.

—The enumeration algorithm picks close to optimal solutions under an hour.

## 6.  RELATED WORK

Anonymizing data consist in a sequence of operations which aims to transform the original table in a way that is suitable for privacy requirements. To achieve privacy, there exist a set of techniques, besides *Generalization and Suppression* seen so far.

*Data perturbation* is a technique that aims to protect information disclosure through modification of either released dataset or microdata information. It implies modification of original data: in such a way the whole sanitized data can be released; otherwise *Output perturbation* refer to modification of a query answer: query is performed over the real dataset, but the results are properly modified to avoid information disclosure. The mean property of such approach is that data collected by data recipient can be considered "synthetic", that is they do not correspond to entity of the real world, but only the statistic selected by the data publisher are preserved during the anonymization process. Data perturbation is widely performed with *Additive noise techniques*: the goal is to replace the original sensitive value $v$ with another value $v + r$ where $r$ is a random value drawn from a distribution. Dwork [Dwork 2006] use noise to achieve strong privacy preserving keeping good accurate data information, through the definition of $\epsilon - privacy$ that captures the increased risk to one's privacy incurred by participating in a database. Another technique of *data perturbation* is *data swapping*. It consists in seeking possible transformation of a dataset, exchanging values of sensitive values such statistical information, frequency counts or other data features are preserved. *Data swapping* approaches can be applied to numerical and categorical attributes; an improved version is named *rank swapping*: values of attribute $A$ are ranked in ascending order, than given a value $v \in A$ it is replaced with another value $u \in A$ randomly chosen within a restricted range having $v$ as central value. There exists also $micro - aggregation$ techniques: data are separated in small groups replacing the value of the chosen attribute with a unique value. For instance, in [D. Defays 1993] $micro - aggregation$ consists in creating groups each one containing three elements and replacing their sensitive attribute with the corresponding mean value.

The goal *data permutation* is to de-associate the relationship between quasi-identifiers and sensitive attribute values: the technique generates a dataset partitioning, then shuffles the values of sensitive attribute of the records within each groups. *Permutation* is based on *Anatomization*, which consists in dividing the dataset into two tables one containing the QIs and the other containing the sensitive values; both table have a common attribute $GroupID$, linking the quasi-identifiers belonging to a group with the sensitive values with the same "GroupID". Both techniques, unlike *generalization*, do not modify the values of quasi-identifier and sensitive attribute.

*Imputation* is another technique that can be used to achieve privacy in data publishing. Surveys can contain several incomplete entries because of different causes: individual does not want to reveal or he simply does not such information. Sometimes data can be correct itself, but in combination with other information, they are not feasible (e.g. a girl who answers that she is fulfilling her military service in a country where the draft does not exists). Comparing datasets containing non-response and response data can lead to leakage of information, thus to privacy breaches. To avoid such problem, the *available method* consist in leave out of the query results all the non-response items, as like they are not selected by the query predicate. This is a non-realistic assumption, though. Three kinds of *imputation* [Nordholt 1998] can be distinguished: *deductive imputation*, where the missing values are deducted by other data, such as *age* can be deducted from *date of birth* and so forth, *deterministic or stochastic imputation*, used if the correct value cannot be deduced. In such a case with *stocastic* method a prediction is made for the non-response value generating random numbers whereas a *deterministic imputation* use values generated in a deterministic way. An example of a deterministic imputation is the imputation of the mean of the known values of some variable for the missing values on that same variable.

This paper is related on *Generalization and Suppression* techniques, which are widely studied in PPDP, illustrating how to incorporate the most important definitions of privacy, as well the latest ones, using a search-based approach for the trade-off privacy-utility.

## 6.1 Other Related Work

Many optimization problems in computer science and related areas can be modelled as a multidimensional partitioning problem, e.g., load balancing in parallel computing applications [Khanna et al. ], histogram construction for selectivity estimation in query optimizers [Poosala et al. 1996; Poosala and Ioannidis 1997], data anonymization for privacy-preserving data publication [Bayardo and Agrawal 2005; LeFevre et al. 2006; Hore et al. 2007] etc. to name a few. In a typical setting, the partitioning problem consists of set of data elements along with an objective function and a set of constraints. The data element are represented as points in some suitably chosen $d$-dimensional space and the optimization goal is to partition this set into a number of bins[11] such that the value of the objective function is minimized (maximized) and at the same time, all the constraints are met. The objective function is often an aggregate function over the resulting bins. For e.g., in the problem of multidimensional histogram creation, if data elements are associated with a weight equal to their frequency in the dataset, the objective is to minimize the total sum of square of deviations from the mean value in each bin. Such a partition tends to make within-bin distribution more uniform and leads to high quality histograms for density estimation problems. The constraints define the subset of partitioning schemes that are admissible (feasible) for the given application. For e.g., space constraints might require the number of total bins to be less than some number or the weight of each bin to be less than some maximum weight bound etc.

---

[11]In many problems the number of bins might not be specified directly, but instead derived from other constraints specific to the application at hand.

Here, we briefly summarize the related previous work and compare and contrast it with ours. The following three examples give a glimpse of the variety of optimization problems that can be modelled as partitioning problems.

**Complex data analysis:** Recently, a framework for summarizing complex datasets was proposed in [Fagin et al. 2005], where the following (new) operations for analyzing multidimensional data were suggested: The operation "divide" computes concise summary of the data set, "differentiate" allows comparison of two different datasets with respect to certain dimension and "discover" breaks up the dataset into groups in order to expose internal patterns. The proposed solutions to all the 3 operations involve partitioning the underlying dataset(s) so as to maximize some quality measure under a given set of constraints. They identify three classes of partitions referred to as "pairwise disjoint classes (PDCs)" which correspond to the generic, special case of hierarchical and grid partitioning schemes shown in figure 1.

**Floorplans for IC design:** In [Yao et al. ] the authors study the problem of generating *floorplans* for VLSI chip design. They specifically look at two kinds of partitions, *hierarchical* and *mosaic* partitions. Mosaic partitions are a subclass of the generic class of rectangular partitions where all intersections form a "T-junction". They show that the set of distinct mosaic plans with $N$ blocks has a one-to-one map with the set of "twin binary trees" where in each tree has $N$ nodes. More details can be found in [Yao et al. ]. It was shown earlier in [Dulucq and Guibert 1998] that the number of distinct pairs of twin binary trees with $N$ nodes is the $N^{th}$ *Baxter number $B(N)$* [Baxter 1964]. Similarly, the authors also show that the the class of 2-dimensional hierarchical rectangular partition (referred to as the "*slicing*" partitions in [Yao et al. ]) can be denoted by the class of *v-h-trees* which were first described in [Szepieniec and Otten 1980]. Similar to mosaic partitioning, the set of hierarchical partitions in 2 dimensions with $n$ blocks is have a one-to-one correspondence with the set of distinct v-h-trees with $n$ leaves. Also, it is shown that the number of distinct hierarchical partitions (floorplans) in 2 dimension is given by twice the *Schroder number $A_n$* [Etherington 1940].

**Anonymization of multidimensional data:** We described some of the existing algorithms for data anonymization in the previous section.

**Approximation algorithms for data partitioning:** Computing optimum rectangular partitions in 2 or higher dimensional space for all of the above optimization problems turns out to be NP-hard. Authors in [Grigni and Manne ] show that it is NP-Hard to partition a $n \times n$ two dimensional array into $p^2$ blocks by partitioning the rows and columns into $p$ intervals each such that the maximum weight of the resulting blocks is minimized. In the general case, each entry of the array is some positive number, but it remains NP-Hard even when the numbers are 0/1 for all entries. Similar optimization problems for hierarchical and generic partitioning schemes have also been shown to be NP-Hard or NP-complete As a result, researchers have focussed on developing approximation algorithms for these problems. For instance, authors in [Khanna et al. 1997] give a $O(1)$ approximate solution to this problem which runs in polynomial time. Authors in [Muthukrish-nan and Suel ] present fast approximation algorithms for a few variants of the above mentioned problem.

## 7. SUMMARY & CONCLUSION

In this paper, we explored the problem of computing optimal $k$-anonymization of multidimensional data using hierarchical partitioning of the space. The approach we took was one of a complete search using systematic enumeration of the solution space. To our knowledge, there is no work that proposes a technique to systematically enumerate all the hierarchical partitioning of a multidimensional space.

In order to make the search for the optimal solution feasible in this huge space of solutions, we employed pruning heuristics to reduce the search space to a manageable size. Our enumeration tree exhibits a nice spatial locality property that allows us to make tight lower bound estimates efficiently by considering local distribution of the data in any region of interest.

Subsequently, we propose a flexible priority queue based algorithm that implements a prioritized search to get to good solutions quickly. In fact, this algorithm can execute in two modes, where one is directed towards finding an optimum solution while the other mode optimizes for finding solutions within a target approximation ratio.

We did extensive experimentation using a variety of combinations of the cost metrics and constraints motivated by some popular measures of privacy and information loss in privacy-preserving data mining applications. In our experiments besides characterizing the performance of our enumeration algorithm, we also carry out exhaustive comparison with some basic greedy heuristics and in the process discover some interesting trends that deviate significantly from what was predicted previously. In the this paper, we report results using the popular DM metric as the cost and k-anonymity, l-diversity and length restrictions as constraints. We also give performance results by varying the priority queue parameters like the priority generating functions and maximum allowed size of the queue.

**Other related work & future directions:** Finally, we note that while this paper is centrally motivated by the need for privacy preservation for data publishing, the problem of generalization based data anonymization is in many ways similar to the problem of optimal histogram construction and other related data partitioning problems. A wide variety of similar optimization problems have been studied earlier in different contexts e.g., query optimization, image compression, parallel computing etc., where similar families of partitionings have been considered [Anily and Federgruen 1991; Muthukrishnan et al. 1997; Berman et al. ; Muthukrishnan and Suel ]. All these applications could also benefit from the current work. In fact, the turn many approximation algorithms proposed in them could perhaps be adapted for the class of anonymization problems we mentioned here. Although we did not pursue these issues in the current work, these remain as some attractive avenues to explore in the future.

ACKNOWLEDGMENTS

acknowledgements

# Constrained Generalization for Data Anonymization: A Systematic Search Based Approach

BIJIT HORE,
University of California, Irvine
RAVI CHANDRA JAMMALAMADAKA,
University of California, Irvine
SHARAD MEHROTRA
University of California, Irvine
AMEDEO D'ASCANIO
University of Bologna, Italy

---

Here, we present the proof of theorem 2 of section 2.2.

THEOREM 2. *Algorithm* **Enumerate(D,S)** *systematically enumerates all distinct hierarchical partitions of the multidimensional space $D$ generated using splits from the given split set $S$.*

Specifically, we prove the following two properties of the enumeration algorithm: (i) **Completeness**: It correctly generates all the distinct hierarchical partitions of the multi-dimensional space using the given set of splits. (ii) **Uniqueness**: It does so without duplication, i.e. there is a one-to-one correspondence between the nodes in partition enumeration tree (PET) and the set of hierarchical partitions of the space. The proof outline is given below.

(1) First, we look at a simplified version of the algorithm **Enumerate** (figure 8) wherein the constraint checking is step (line 3) is simplified to only check compliance with Constraint 1. We call this the **MultiEnumerate** algorithm (figure 25). That is, it generates a multi-way tree of all timestamped $KD$-trees constructible using the given set of splits which comply with Constraint 1 (i.e. left subspace of a $KD$-tree node has been partitioned before the right subspace). Recall that a single hierarchical partition may be denoted by several different $KD$-trees. Since the simplified algorithm generates all $KD$-trees representing the same partition, we call it the **MultiEnumerate** algorithm. Therefore, the set of hierarchical partitions induce an equivalence relation over this set of $KD$-trees, where each equivalence class represents a distinct hierarchical partition of the space. We will use the method of induction to prove the following claim:

---

> CLAIM .1. *Algorithm* **MultiEnumerate** *enumerates all members of each equivalence class, therefore ensuring completeness.*

(2) Second, we look the effect of introducing the check for compliance with Constraint 2 into algorithm **MultiEnumerate**. The "Detect-Legal-Split" (DLS) routine in algorithm **Enumerate** (figure 8) first checks for Constraint 1 which is already implemented by **MultiEnumerate**. Therefore we only concentrate on the effect of enforcing compliance with Constraint 2 by the DLS routine. We explain in detail, how avoiding "out-of-sequence splits" removes all duplicates and retains exactly one "copy" of each hierarchical partition. Specifically we will prove the following claim.

> CLAIM .2. *DLS returns TRUE for exactly one member from each equivalence class induced by the set of hierarchical partitions. This unique $KD$-tree is the only "legal" representative of the hierarchical partition corresponding to the class (this $KD$-tree is called the "partition tree" for that partition).*

(3) Lastly, we make the following observation.

> **Observation** 2. *Algorithm* **Enumerate** *generates new partitions by adding a single split at a time to an existing partition. As a result, legal $KD$-trees (corresponding to child nodes) can only be derived from a parent $KD$-tree that is legal (i.e. complies with Constrain 1 and 2).*

From the above observation, we are guaranteed completeness by simply retaining legal $KD$-trees at any stage in the enumeration process. This allows us to safely drop all trees derived from a legal $KD$-tree that do not meet the DLS criteria.

Combining the three arguments made above, it follows that algorithm **Enumerate** (in figure 8) generates all distinct hierarchical partitions of the space using the set of splits and does so without any duplication. We now prove the claims in item 1 and 2 above, the claim in item 3 is clear from the arguments provided there.

### .1 Proof of Claim 1

The **MultiEnumerate** algorithm is given below in figure 25. We use induction on $n$, cardinality of the set $Seq_{[n]}$ of splits and prove that **MultiEnumerate** generates the complete set of $KD$-trees denoting the set of hierarchical partitions.

Note that we only need to prove the correctness for those instances of **MultiEnumerate** where the input parameter $T(r)$ denotes a degenerate $KD$-tree, i.e. $T(r)$ denotes an undivided subspace. We will denote such a tree by $T_\phi$. Making the hypothesis for the more general case where $T(r)$ can be any $KD$-tree unnecessarily complicates the proof procedure, hence we make the simpler hypothesis which suffices. The proof by induction is given below.

For $n = 1$, the algorithm generates the enumeration tree containing two nodes, one denoting the undivided space and the other denoting the partition having just one split that dividing the space into two subspaces. Now we make the induction hypothesis for all values of $n$ from 1 to $k$:

**Induction hypothesis**: For all values of $n = 1, \ldots, k$ Algorithm **MultiEnumerate** $(D, Seq_{[n]}, r, T_\phi)$ generates the enumeration tree rooted at $r$, consisting of all distinct timestamped $KD$-trees that are constructible on the domain space of data set $D$, using zero or more splits from the set $Seq_{[n]}$. Additionally each tree generated complies with Constraint 1.

---

**MultiEnumerate(D, Seq, r, T(r))**
**Output:** Construct the enumeration tree
        (rooted at $r$) of all $KD$-trees
        extending $T(r)$ and comply with
        Constraint 1
**BEGIN**
1)  **For Each** leaf $l \in T(r)$ s.t. $l \models$ Constraint 1
2)    **For Each** available split $s_l \in Seq$ at $l$
3)      Generate new child node $c$
4)      Generate $KD$-tree $T(c) \leftarrow T(r) \cup s_l$
5)      Add pointers $r \rightarrow c$ ; $c \rightarrow r$
6)      **MultiEnumerate**$(D, Seq, c, T(c))$
7)    **End For**
8)  **End For**
9)  **Return** $r$
**END**

---

Fig. 25.   Generating the complete set of $KD$-trees using $n$ splits

---

**InductiveEnumerate(D, Seq$_{[k+1]}$)**
**Output**: The set $TREES$ of all $KD$-trees
**BEGIN**
1)   $TREES \leftarrow \{T_\phi\}$
2)  **For Each** split $s_i \in Seq_{[k+1]}$
3)    $\mathfrak{T}_{left} \leftarrow$ **InductiveEnumerate**
            $(D_{left(s_i)}, Seq_{[k+1]} - \{s_i\})$
4)    $\mathfrak{T}_{right} \leftarrow$ **InductiveEnumerate**
            $(D_{right(s_i)}, Seq_{[k+1]} - \{s_i\})$
5)    **For Each** $T^l \in \mathfrak{T}_{left}$
6)      Add 1 to the timestamp of each node in $T^l$
7)      **For Each** $T^r \in \mathfrak{T}_{right}$
8)        Add **MAX** $\{timestamp(node \in T^l)\}$ to
           timestamp of each node in $T^r$
9)        Generate $T_{new} \leftarrow T_\phi$
10)      $T_{new}.root.split \leftarrow (s_i)$
11)      $T_{new}.root.timestamp \leftarrow [1]$
12)      $T_{new}.root.leftchild \leftarrow T^l$
13)      $T_{new}.root.rightchild \leftarrow T^r$
14)      $TREES \leftarrow TREES \cup \{T_{new}\}$
15)      **End For**
16)    **End For**
17)  **End For**
18)  **Return** $TREES$
**END**

---

Fig. 26.   Re-writing MultiEnumerate for inductive proof

**Induction step:** To prove that **MultiEnumerate** generates the complete set of timestamped $KD$-trees when called with any split set of size $k + 1$.

**Proof of induction step**: Consider the following instance of the algorithm: **MultiEnumerate** $(D, Seq_{[k+1]}, r, T_\phi)$. Any such instance of the **MultiEnumerate** algorithm can be re-written in a non-constructive but equivalent form as the new algorithm **InductiveEnumerate** shown in figure 26. Note that the two algorithms are completely equivalent for the class of instances of **MultiEnumerate** that we are interested in (i.e. where $T(r) = T_\phi$ always). Note that equivalent algorithm does not require the inputs $r$ and $T(r)$ and outputs the complete set of $KD$-trees instead of an enumeration tree (hence we call it non-constructive). The key observation is that the split that is chosen first cuts across the whole space, therefore it cannot be used again and hence can be dropped from the set of available splits in both the recursive calls to the function. The algorithm introduces the first split and calls itself recursively on the space to the left and right of the first split with a split set of size one less than the one in its input. Therefore the proof follows from our inductive hypothesis and the equivalence of the two algorithms **MultiEnumerate** and **InductiveEnumerate**.

### .2 Proof of Claim 2

From here onwards, we will assume that all $KD$-trees comply with Constraint 1 unless otherwise stated. The first part of the DLS routine checks compliance with Constraint 1 and since that is assumed to be taken care of, the phrase "DLS returns true" will be used to imply compliance with Constraint 2 (i.e. splits at nodes are in-sequence or not).

**Nature of DLS routine and the enumeration process**: Observe that the DLS routine, invoked at any node $t$ of a $KD$-tree $T$ checks whether split $s(t)$ at $t$ is out-of-sequence with the split $s(a)$ at any ancestor $a$ of $t$. $T$ is considered legal (i.e. $T$ is a partition tree) if and only if the DLS routine returns true for each node in $T$. Since algorithm **Enumerate** generates new trees by extending existing partition trees by splitting a single partition block at a time, it is only necessary to invoke DLS once for every new partition generated from the parent node in the enumeration tree. This also implies that there can be no partition tree (i.e. legal $KD$-tree) that is derived from an "illegal" parent $KD$-tree and therefore only legal $KD$-trees need be generated and retained during the entire enumeration process.

**Parent-child switching of splits**: We now introduce an operation called "parent-child switch" which can be performed between a node and its children (or a child, it will become clear soon) in a $KD$-tree under certain conditions. The important property of such an operation is that, they generate a new $KD$-tree that denotes the same partition, but simply changes the sequence in which the cutting splits are introduced in some subspace. We give the definition below and provide an example to illustrate the switch operation.

**Definition** 13. **Parent-child switch**: *There are two distinct cases in which a switch can be made:*

*Case 1) In a $KD$-tree $T$, a node $p$ and its two children $c_{left}$ and $c_{right}$ have node splits such that $s(c_{left}) = s(c_{right}) = s_2 \neq s(p) = s_1$, then a parent-child switch*

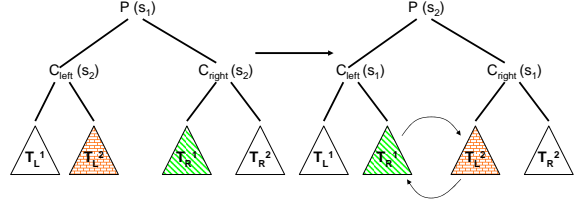

Fig. 27.   Parent-child switch for orthogonal splits (case 1)

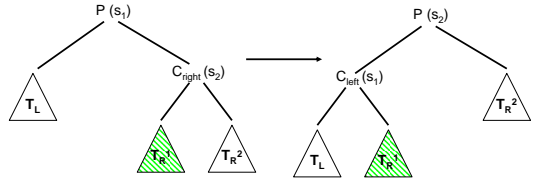

Fig. 28.   Parent-child switch for parallel splits (case 2)

*(or simply a switch) refers to the atomic operation that switches the positions of splits $s_1$ with $s_2$, followed by switching the positions of the right subtree of $c_{left}$ with the left subtree of $c_{right}$. Note that visualizing geometrically, this case holds when the two splits are orthogonal to each other (i.e. along two different dimensions (attributes)). This case is illustrated in figure 27.*

**Definition** 14. *Case 2) This corresponds to splits $s_1$ and $s_2$ that are parallel in space. Say $s(p) = s_1 \neq s(c_{right}) = s_2$ (when $s_2$ is a split to the right of $s_1$ in the space), then a switch refers to the atomic operation that "left rotates" the nodes $p$ and $c_{right}$. That is, $c_{right}$ becomes the parent with split $s_2$, $p$ becomes the left child of $c_{right}$ and positions of the left subtree of $c_{right}$ is switched with that of the right subtree of $p$. The "right rotate" operations is symmetric. This case is illustrated in figure 28.*

**Two part proof**: We give a constructive proof and show that each equivalence class of $KD$-trees always has exactly one tree that satisfies the DLS routine at all of its nodes. The proof has two part: 1) We show that for any $KD$-tree with an out-of-sequence pair of splits, say $< (s), (s') >$ such that $(s)$ is the split introduced earlier, we can always derive a new tree using a sequence of parent-child switch operations such that it represents the same hierarchical partition and has $(s')$ as the earlier of the two splits. Specifically we will show using straight forward induction that all conflicts (i.e. pairs of out-of-sequence splits) can be resolved for any given $KD$-tree, thereby generating a conflict-free $KD$-tree. 2) We show there can be at most one such tree for every equivalence class.

**Proof of the first part**: The proof is based on the following three observations, first of which we have already stated above.

**Observation** 3. *Any parent-child switch operation is "partition preserving".*

*That is, the KD-tree before and after a switch is made, represents the same hierarchical partition.*

**Observation** 4. *A parent-child switch operation between the node p and its child node(s) never generates a new conflict with any ancestor of p, i.e. never generates new out-of-sequence pairs with split at an ancestor[12] of p.*

**Observation** 5. *If $s^*$ is the split at a KD-tree node n that forms a cut of the subspace denoted by some ancestor a of n, then $s^*$ is also a cut across the subspace denoted by each node on the path $a \rightsquigarrow n$.*

Now look at the **MultiEnumerate** algorithm, in each iteration it starts with a legal KD-tree (i.e. conflict free) in the enumeration tree, call it the parent tree and generates a new candidate tree by adding a single split to some leaf of the parent tree. One only needs to check if there is a conflict between this split added at the newly formed node and the split at each of its ancestor nodes. We use induction as mentioned above to show that a conflict free tree exists for the given partition.

It is easy to see that when the parent tree has one node (i.e. one root node and two leaves), a conflicting split at a newly split node can be resolved using a switch operations with the root. Let us assume that all conflicts can be resolved for a KD-tree with up to $k$ nodes. Therefore we can always have a legal KD-tree with $k$ nodes. Now say a new node $n_{new}$ is generated thereby making it a tree with $k+1$ nodes. Now if the highest ancestor $a$ of this node, with which it conflicts is a node other than the root (i.e. $a \neq root$), by induction hypothesis we can always resolve the conflict in the subtree rooted at $a$ since it is a tree with strictly $\leq k$ nodes. From observation 4, the conflict can be resolved without introducing any new conflicts with any ancestor of $a$. Therefore there exists a legal KD-tree representing the new partition. Else if, the new node conflicts with the root of the tree, using observations 3, 4 and 5, we can always resolve the conflict with the root node, which will leave us with at most two "conflict-prone" subtrees (left and right subtree of root node), which can again be made conflict free using our induction hypothesis. Hence the first part is proved.

**Proof of the second part**: Assume we are given two randomly chosen KD-trees $T_1$ and $T_2$ belonging to the same equivalence class. Now, using parent-child switch operations let $T_1$ and $T_2$ be transformed into the conflict-free trees $T_1^*$ and $T_2^*$ respectively. Now, since these trees represent the same hierarchical partitions, they will have the same number of leaves and in fact there is a 1-to-1 correspondence of the set of leaves in $T_1^*$ to those in $T_2^*$. Let us assume that these two final trees are distinct, then we can find at least one pair of leaves (corresponding to the same partition block in each tree) such that the sequence of splits along the root-to-leaf path in one tree is different from the other (see figure 29). Let us call these paths $path_1$ and $path_2$. Since these two are not the same, let $n_1^*$ and $n_2^*$ be the first nodes in each path, where the splits do not match, say denoted by $(s_1^*)$ and $(s_2^*)$. But both $n_1^*$ and $n_2^*$ denote the same subspace and the splits chosen at these nodes are cuts of this subspace. And now, since one of the two splits $(s_1^*)$ or $(s_2^*)$ has a higher priority than the other, only the path with a node-split of a higher priority could be a valid path in a conflict-free tree. Thereby contradicting our initial assumption that $T_1^*$ and $T_2^*$ are distinct. Hence we prove that at most one conflict free tree is there for each equivalence class.

---

[12]A switch operation may generate new conflicting pairs in either or both of the subtrees rooted at $p$.

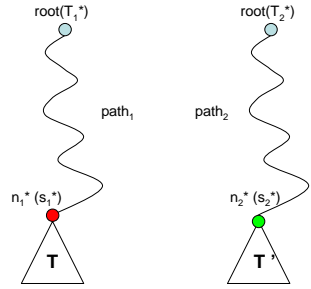

Fig. 29. Uniqueness of the conflict-free $KD$-tree

Now the proof of Claim 2 is complete.